\documentclass[useAMS,usenatbib]{mn2e}

\usepackage{epsfig,graphicx,mathptm}
\usepackage{times}%
\usepackage{natbib}

\newcommand{\gtrsim}{\ga}
\newcommand{\lesssim}{\la} 

\newcommand{\sbunits}{erg s$^{-1}$ cm$^{-2}$ deg$^{-2}$}

\newcommand{\ovisc}{{\it ovisc}}
\newcommand{\lvisc}{{\it lvisc}}
\newcommand{\csf}{{\it csf}}
\newcommand{\csfc}{{\it csfc}}

\def\aap{A\&A}
\def\apj{ApJ}

\def\apjl{ApJ}
\def\mnras{MNRAS}

\def\aaps{A\&A Supp.}

\title[]
{Simulated X-ray galaxy clusters at the virial radius: slopes of the 
gas density, temperature and surface brightness profiles}

\author[M. Roncarelli et al.]
{\parbox[]{7.in}
{M. Roncarelli$^1$, 
S. Ettori$^2$, 
K. Dolag$^3$, 
L. Moscardini$^{1,4,5}$, 
S. Borgani$^{6,4,5}$, 
G. Murante$^7$
\\~\\
\footnotesize
$^1$ Dipartimento di Astronomia, Universit\`a di Bologna, via Ranzani
  1, I-40127 Bologna, Italy (mauro.roncarelli, lauro.moscardini@unibo.it)\\
$^2$ INAF, Osservatorio Astronomico di Bologna, via Ranzani
  1, I-40127 Bologna, Italy (stefano.ettori@oabo.inaf.it) \\
$^3$ Max-Planck-Institut f\"ur Astrophysik, Karl-Schwarzschild Strasse
  1, D-85741 Garching bei M\"unchen, Germany (kdolag@mpa-garching.mpg.de)\\
$^4$ INAF -- National Institute for Astrophysics, Italy \\
$^5$ INFN -- National Institute for Nuclear Physics, Italy \\
$^6$ Dipartimento di Astronomia dell'Universit\`a di Trieste, via
  Tiepolo 11, I-34131 Trieste, Italy (borgani@oats.inaf.it)\\
$^7$ INAF, Osservatorio Astronomico di Torino, strada Osservatorio 
  20, I-10025 Pino Torinese (TO), Italy (giuseppe@to.astro.it)
}}

\begin{document}


\pagerange{\pageref{firstpage}--\pageref{lastpage}} \pubyear{2006}

\maketitle

\label{firstpage}

\begin{abstract}
Using a set of hydrodynamical simulations of 9 galaxy clusters with masses 
in the range $1.5 \times 10^{14} M_{\odot} < M_{\rm vir} < 3.4 \times 10^{15} M_{\odot}$, 
we have studied the density, temperature and X-ray surface brightness
profiles of the intra-cluster medium in the regions around the virial radius.
We have analyzed the profiles in the radial range well above the cluster core,
the physics of which are still unclear and matter of tension between 
simulated and observed properties, and up to the virial radius and beyond,
where present observations are unable to provide any constraints.
We have modeled the radial profiles between $0.3 R_{200}$ and $3 R_{200}$
with power laws with one index, two indexes and a rolling index.
The simulated temperature and [0.5--2] keV surface brightness profiles
well reproduce the observed behaviours outside the core.
The shape of all these profiles in the radial range considered depends
mainly on the activity of the gravitational collapse, with no
significant difference among models including extra-physics.
The profiles steepen in the outskirts, with the slope of the power-law
fit that changes from -2.5 to -3.4 in the gas density, 
from -0.5 to -1.8 in the gas temperature, 
and from -3.5 to -5.0 in the X-ray soft surface brightness.
We predict that the gas density, temperature and [0.5--2] keV surface brightness
values at $R_{200}$ are, on average, $0.05$, $0.60$, $0.008$ times the measured
values at $0.3 R_{200}$. At $2 R_{200}$, these values decrease by an order
of magnitude in the gas density and surface brightness, 
by a factor of 2 in the temperature, putting stringent limits on the 
detectable properties of the intracluster-medium (ICM) in the virial regions.
\end{abstract}


\section{Introduction} \label{sect:intro}

Galaxy clusters form in correspondence of the highest peaks 
of the fluctuations in the cosmic dark matter density field and
collapse under the action of gravitational attraction
to define structures with a mean density enhanced with respect
to the cosmic value by a factor of few hundreds.
They reach the virial equilibrium over a volume with a typical
virial radius that indicates, as a first approximation, 
the regions where the pristine gas accretes on the dark matter 
halo through gravitational collapse and is heated up
to millions degrees through adiabatic compression and
shocks. 
This intracluster-medium (ICM), that represents about 80 (15) per cent
of the baryonic (total) cluster mass \citep[e.g.][]{ettori2004}, 
becomes, then, X-ray emitter mainly through bremsstrahlung
processes, thus allowing one to trace and characterize the distribution 
of the baryonic and dark matter components.

From an observational point of view, the measurement of the
properties of the ICM is possible only where the X-ray emission can 
be well resolved against the background (both instrumental and cosmic).
At the present, data permit to firmly characterize observables
like surface brightness \citep[see][]{mohr1999, ettori1999} and gas 
temperature \citep[see][]{markevitch1998, degrandi2002, pointecouteau2005, 
vikhlinin2006, zhang2006} out to a fraction ($\sim 0.5-0.6$) of the virial radius.
Only few examples of nearby X-ray bright clusters with surface brightness 
estimated out to the virial radius are available, thanks to 
the good spatial resolution, the large field-of-view and the low 
instrumental background of the 
{\it ROSAT/PSPC} instrument \citep{vikhlinin1999, neumann2005}.
However, since the regions not-yet resolved occupy 
almost 80 per cent of the total volume, they retain most of the 
information on the processes that characterize the accretion
and evolution within the cluster of the main baryonic 
component \citep{molendi2004}.

\begin{table*}
\begin{center}
\caption{
List of main physical parameters of the 9 clusters taken from the \ovisc\
runs. The objects are divided into 2
subsamples according to their virial mass. Note that $T_{\rm vir}$ is the
average mass-weighted temperature of the gas up to $R_{\rm vir}$ while
$T_{200}$ is the temperature of the gas at $R_{200}$.}
\begin{tabular}{lccccccccccc}
 & \multicolumn{4}{c}{Sample $A$: $M_{\rm vir}>10^{15} M_\odot$} &  &  &
\multicolumn{5}{c}{Sample $B$: $M_{\rm vir}<10^{15} M_\odot$} \\
Cluster name & {\it g1} & {\it g8} & {\it g51} & {\it g72} &  &  & {\it g676} & {\it g914} & {\it
g1542} & {\it g3344} & {\it g6212} \\
\hline
$M_{\rm vir}$ ($10^{15} M_\odot$) & 2.12 & 3.39 & 1.90 & 1.96 &  & &
                                    0.15 & 0.17 & 0.15 & 0.16 & 0.16  \\

$R_{\rm vir}$ (Mpc)               & 3.37 & 3.94 & 3.25 & 3.28 &  & &
                                    1.40 & 1.46 & 1.40 & 1.43 & 1.43  \\

$T_{\rm vir}$ (keV)               & 6.49 & 9.89 & 5.78 & 5.44 &  & &
                                    1.25 & 1.28 & 1.18 & 1.26 & 1.22  \\

\hline
$R_{200}$ (Mpc)                   & 2.48 & 2.91 & 2.39 & 2.40 &  & &
                                    1.04 & 1.08 & 1.04 & 1.05 & 1.05  \\

$T_{200}$ (keV)                & 4.18 & 7.52 & 4.18 & 4.09 &  & &
                                    0.82 & 1.03 & 0.83 & 0.88 & 0.84  \\
\hline
\label{tab:9clusters}
\end{tabular}
\end{center}
\end{table*}

\begin{figure*}
\includegraphics[width=0.78\textwidth]{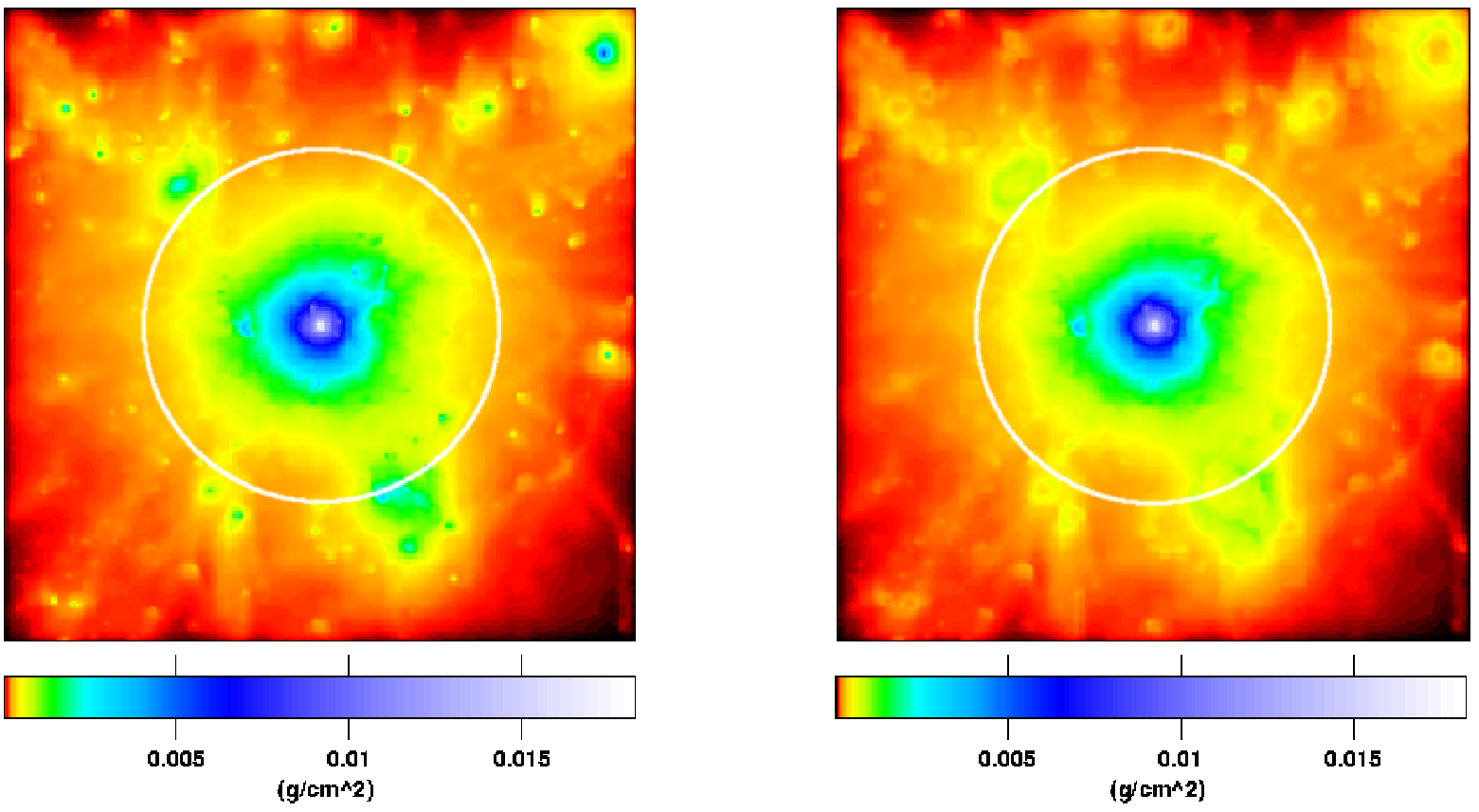}
\includegraphics[width=0.78\textwidth]{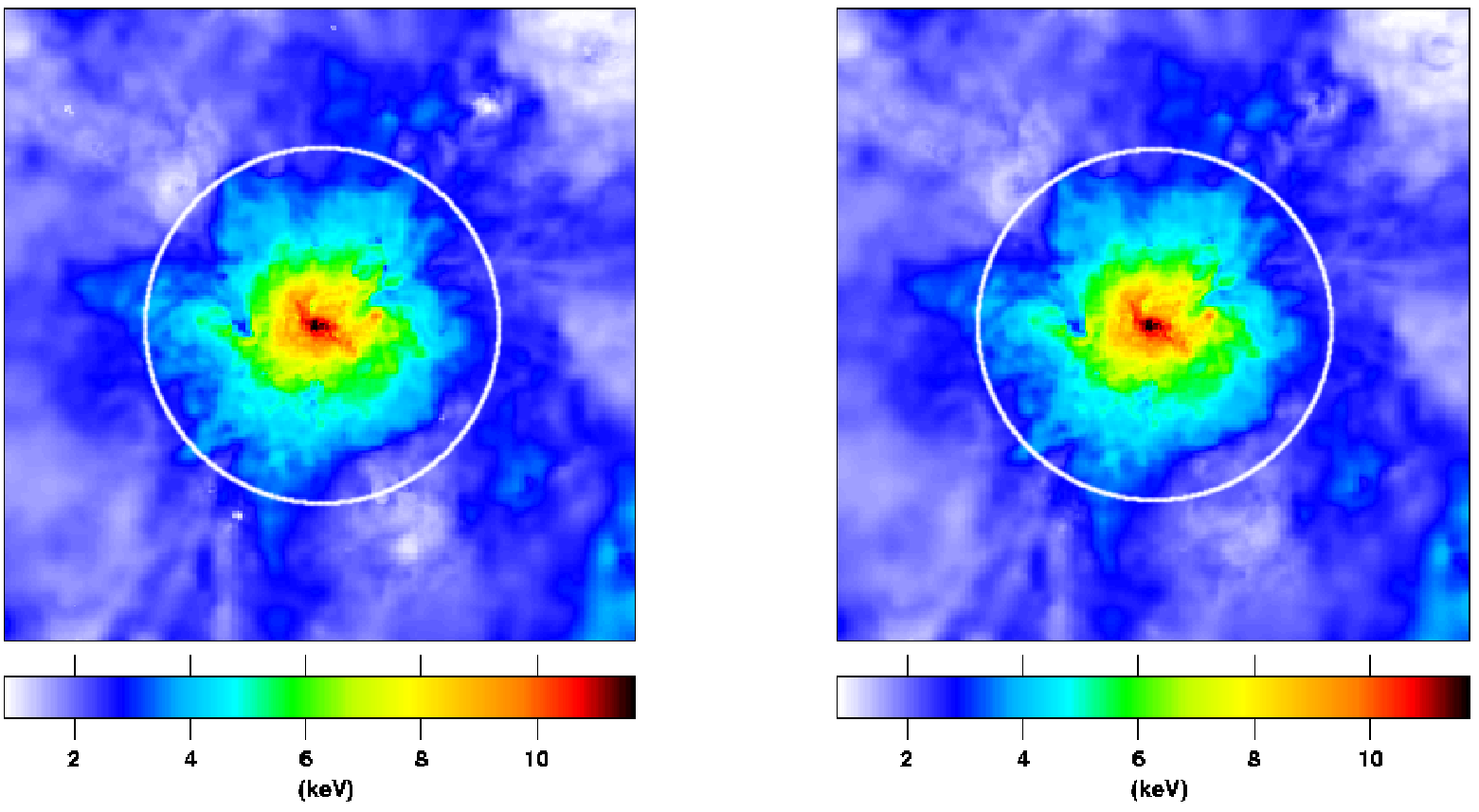}
\includegraphics[width=0.78\textwidth]{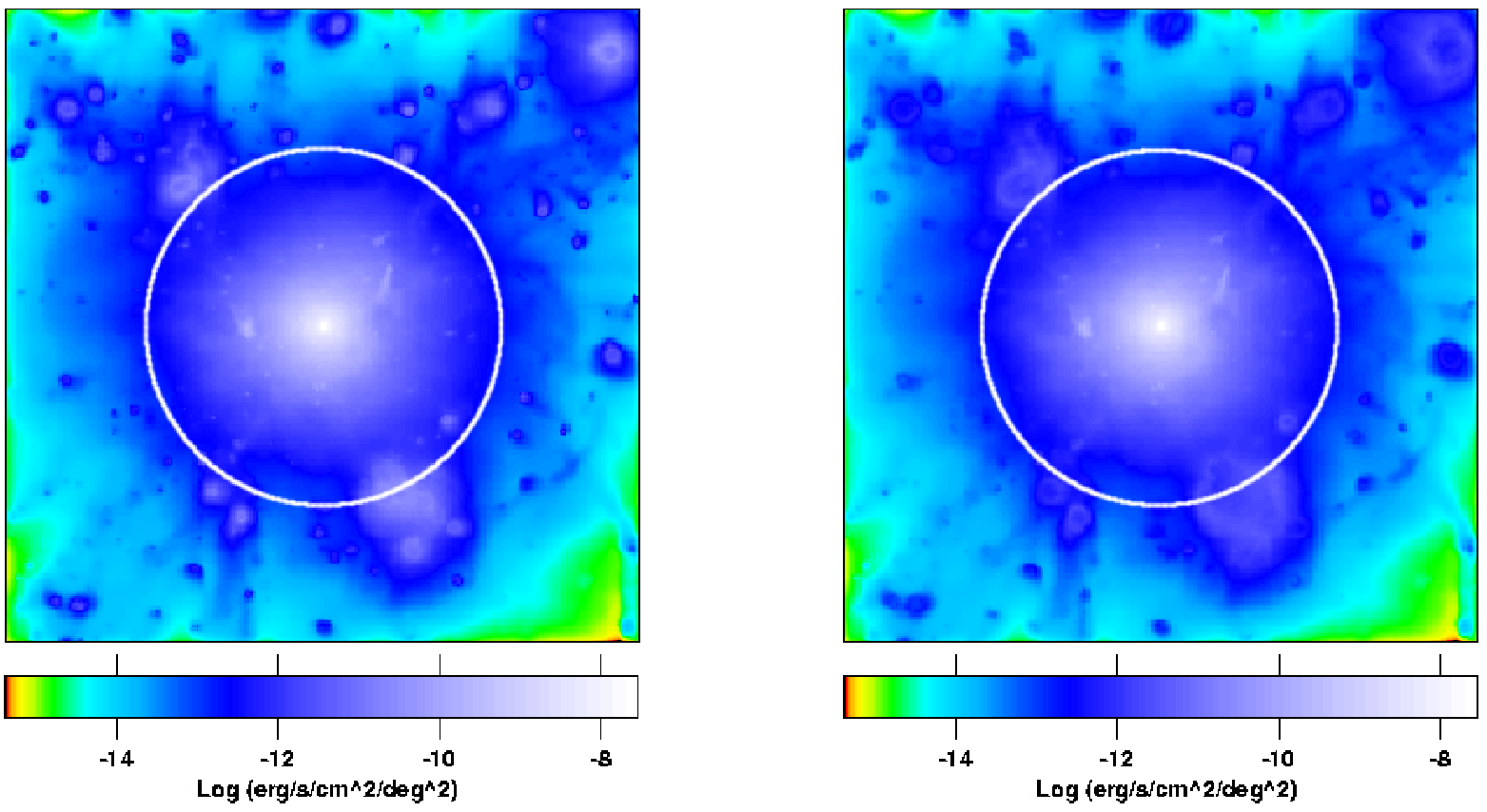}
\caption{
From top to bottom: maps of the projected gas density, mass-weighted temperature 
and soft (0.5-2 keV) X-ray emission of the {\it g1} cluster (\csf\ model)
for the 100 per cent (left) and 95 per cent (right) volume-selection scheme.
The circles indicate the virial radius. The size of the side of each map is 12 Mpc, 
so they cover roughly up to $2.5 R_{200}$.
} \label{fig:maps}
\end{figure*}

In the present work, we study the distribution of the ICM density,
temperature and surface brightness in a set of simulated
galaxy clusters by looking at their radial profiles
in the range between $0.3 - 3 \times R_{200}$\footnote{
In this paper we define the typical physical quantities of clusters as follows:
$R_{200}$ ($R_{500}$) is the radius of a sphere
centered in a local minimum of the potential and enclosing an average
density of $\rho = 200 (500) \rho_{\rm cr}$, 
with $\rho_{\rm cr}$ being the critical density of the Universe. 
The virial quantities are defined using
the density threshold obtained by the spherical collapse model
with the cosmological parameters of the simulations. So the
virial radius $R_{\rm vir}$ encloses an average
density of $\rho_{\rm vir} \simeq 104 \rho_{\rm cr}$ \citep[see][eq.~4]{eke1998}. 
Consequently the virial mass
$M_{\rm vir}$ is the total mass (DM plus baryons) of the cluster up  to
$R_{\rm vir}$.}.
We avoid considering the regions where the clusters evolve a core 
with active cooling and suspected feedback regulation, 
being this balance still matter of tension between the present 
simulated constraints and the observed counterparts
\citep[e.g.][]{borgani2004}.
Therefore, we concentrate our analysis on cluster volumes where
the gas density and temperature suggest that radiative processes
are not dominant and the main physical driver is just the gravitational
collapse, that is expected to be well described 
in the present numerical N-body simulations.
We consider objects extracted from a dark matter only cosmological simulation
that were resimulated at much higher resolution by including the gas subjected
to gravitational heating and other physical treatments like cooling, star
formation, feedback, thermal conduction and an alternative implementation of
artificial viscosity.

We organize this paper in the following way: in the next Section, we
describe our dataset of simulated clusters and their different physical properties;
in Section \ref{sect:results} we discuss the results on the slopes of the gas temperature, 
density and surface brightness profiles in the outskirts 
(i.e. $r > 0.3 R_{200}$) of the simulated objects and compare
our results with the present constraints obtained from X-ray 
observations of bright clusters.
We summarize and discuss our results in Section \ref{sect:concl}.

\section{The simulated clusters} \label{sect:simul}

For the purpose of this work we use a set of 9 resimulated 
galaxy clusters extracted from a dark matter (DM)-only simulation
of a ``concordance'' $\Lambda$CDM model with $\Omega_{\rm m}=0.3$ and 
$\Omega_{\Lambda}=0.7$, using a Hubble parameter $h=0.7$ (the 
Hubble constant being $H_0=h\ 100$ km s$^{-1}$Mpc$^{-1}$) and a baryon 
fraction of $\Omega_{\rm b}=0.019 h^{-2}$. The initial conditions 
of the parent simulation, that followed the evolution of a comoving 
box of $479 h^{-1}$ Mpc per size, were set considering a 
cold dark matter power spectrum normalized by assuming 
$\sigma_8=0.9$ \citep[see][]{yoshida2001}. Our simulations were carried 
out using {\tt GADGET-2} 
\citep{springel2005}, a new version of the Tree-Smoothing Particles 
Hydrodynamics (SPH) parallel code 
{\tt GADGET} \citep{springel2001} which includes an entropy-conserving 
formulation of SPH \citep{springel2002} and a treatment of many 
physical processes affecting the baryon component that can be turned 
on and off to study their influence on the thermodynamics of the gas.
In the volume of the parent simulation 9 different haloes were identified, 
spanning different mass ranges:
4 of them with the typical dimension of clusters and 5 of small clusters or 
group-like objects. Then using the ''Zoomed Initial Conditions'' (ZIC) 
technique \citep{tormen1997} they were resimulated by identifying their 
corresponding Lagrangian regions in the initial domain and populating them 
with more particles (of both DM and gas), while appropriately adding 
high-frequency modes. At the same time the volume outside the region of interest 
was resimulated using low-resolution (LR) particles in order to follow the
tidal effects of the cosmological environment. 
The setup of initial conditions of all the 
resimulations was optimized to guarantee a volume around the clusters of $\sim 
5 R_{\rm vir}$ free of contamination from LR particles. This 
was obtained using an iterative process as follows. Starting from a first guess 
of the high-resolution (HR) region that we want to resimulate, we 
run a DM-only re-simulation. Analyzing its final 
output, we identify all the particles that are at distances smaller 
than 5 $R_{\rm vir}$ from the cluster centre in order to identify 
the corresponding Lagrangian region in the initial domain. Applying ZIC, 
we generate new initial conditions at higher resolution and run one more 
DM-only resimulation.
This procedure is iteratively repeated until we find that none of the LR 
particles enters the HR region, which could be possible because of 
the introduction of low-scale modes. Thus we can safely say that these 
resimulations are representing the external regions of our cluster 
without any spurious numerical effect.

These resimulations are set to have a mass resolution of 
$m_{\rm DM}= 1.13 \times 10^9 h^{-1} 
M_\odot$ for DM particles and $m_{\rm gas}= 1.69 \times 10^8 h^{-1} 
M_\odot$ for baryons, so that every resimulated cluster is resolved with between 
$2 \times 10^5$ and $4 \times 10^6$ particles (both DM and gas) depending on its 
final mass. All the simulations have a (Plummer-equivalent) softening-length 
kept fixed at $\varepsilon = 30 h^{-1}$ kpc comoving at $z > 5$ and switched to a 
physical softening length $\epsilon = 5 h^{-1}$ kpc at lower redshifts.

In order to study the impact of different physical processes on the 
clusters properties, we simulated every cluster using 4 different 
sets of physical models:
\begin{enumerate}

\item Gravitational heating only with an ordinary treatment of gas viscosity 
      (which will be referred as the \ovisc\ model).

\item Gravitational heating only but with an alternative 
      implementation of artificial viscosity (\lvisc\ model), 
      following the scheme proposed by \cite{morris1997}, in which 
      every gas particle evolves with its own viscosity parameter. 
      With this implementation the shocks produced by gas accretion 
      are as well captured as in the \ovisc\ model, 
      while regions with no shocks are not characterized by 
      a large residual numerical viscosity. As already studied in 
      \cite{dolag2005b}, this scheme allows to better resolve the 
      turbulence driven by fluid instabilities, thus allowing 
      clusters to build up a sufficient level of turbulence-powered 
      instabilities along the surfaces of the large-scale velocity 
      structures present in cosmic structure formation.
      
\item Runs which implement a treatment of cooling, star formation
      and feedback (\csf\ model). The star formation is followed by adopting 
      a sub-resolution multiphase model for the interstellar medium
      which includes also the feedback from supernovae (SN) and 
      galactic outflows \citep{springel2003}. The efficiency of SN 
      to power galactic winds has been set to 50 per cent, which turns 
      into a wind speed of 340 km/s.

\item Runs like \csf\ but also including thermal conduction 
      (\csfc\ model), adopting the scheme described by \cite{jubelgas2004}.
      This implementation in SPH, which has been proved to be stable 
      and to conserve the thermal energy even when individual and 
      adaptive time-steps are used, assumes an isotropic effective 
      conductivity parametrized as a fixed fraction of the Spitzer 
      rate (1/3 in our simulations). It also accounts for saturation 
      which can become relevant in low-density regions. For more 
      details on the properties of simulated clusters using this model,
      see also \cite{dolag2004}.

\end{enumerate}

The masses, radii and temperatures\footnote{
Notice that in this paper we prefer to
quote all the temperatures by using the mass-weighted estimator because 
it is more
related to the energetics involved in the process of structure
formation. As shown in earlier papers
\citep[see, e.g.][]{mazzotta2004,gardini2004,rasia2005}, the
application of the emission-weighted temperature, even if it was
originally introduced to extract from hydrodynamical simulations
values directly comparable to the observational spectroscopic
measurements, introduces systematic biases when the structures are
thermally complex. We verified that the use of the 
spectroscopic-like temperature of \cite{mazzotta2004} does not produce 
significant changes in our results for $k_B T >$1 keV.} 
of the 9 clusters for the \ovisc\ model are summarized in Table 
\ref{tab:9clusters}. The other models have similar results for 
masses and radii with variation of few percent between the models while 
temperatures have larger variations with \csf\ and \csfc\  
models having slightly higher temperatures ($\sim 8$ per cent).
In our following analyses we 
will separate the clusters into two different samples: sample $A$ (4 clusters) 
and $B$ (5) according to their virial mass being larger or smaller than 
$10^{15} M_\odot$, respectively.

When considering the environmental properties of these haloes, we must point out 
that the 4 objects of sample $A$ are among the most massive clusters of the volume of the 
parent simulation while the 5 groups were chosen at random with the only 
 condition to be far away from the 4 clusters. At the end of the resimulations 
all the haloes of sample $B$ appear to be \emph{isolated}, e.g. there is no 
halo with mass $M > 10^{13.5} h^{-1} M_\odot$ at distance lower than $5 R_{\rm vir}$; 
on the contrary the big clusters have several small structures inside the volume 
of their resimulation and all of them underwent significant merging activities.
As we will see in Section \ref{sect:results}, this aspect is important to understand the 
different properties of the X-ray profiles of the two samples.

\begin{figure*}
\includegraphics[width=0.43\textwidth]{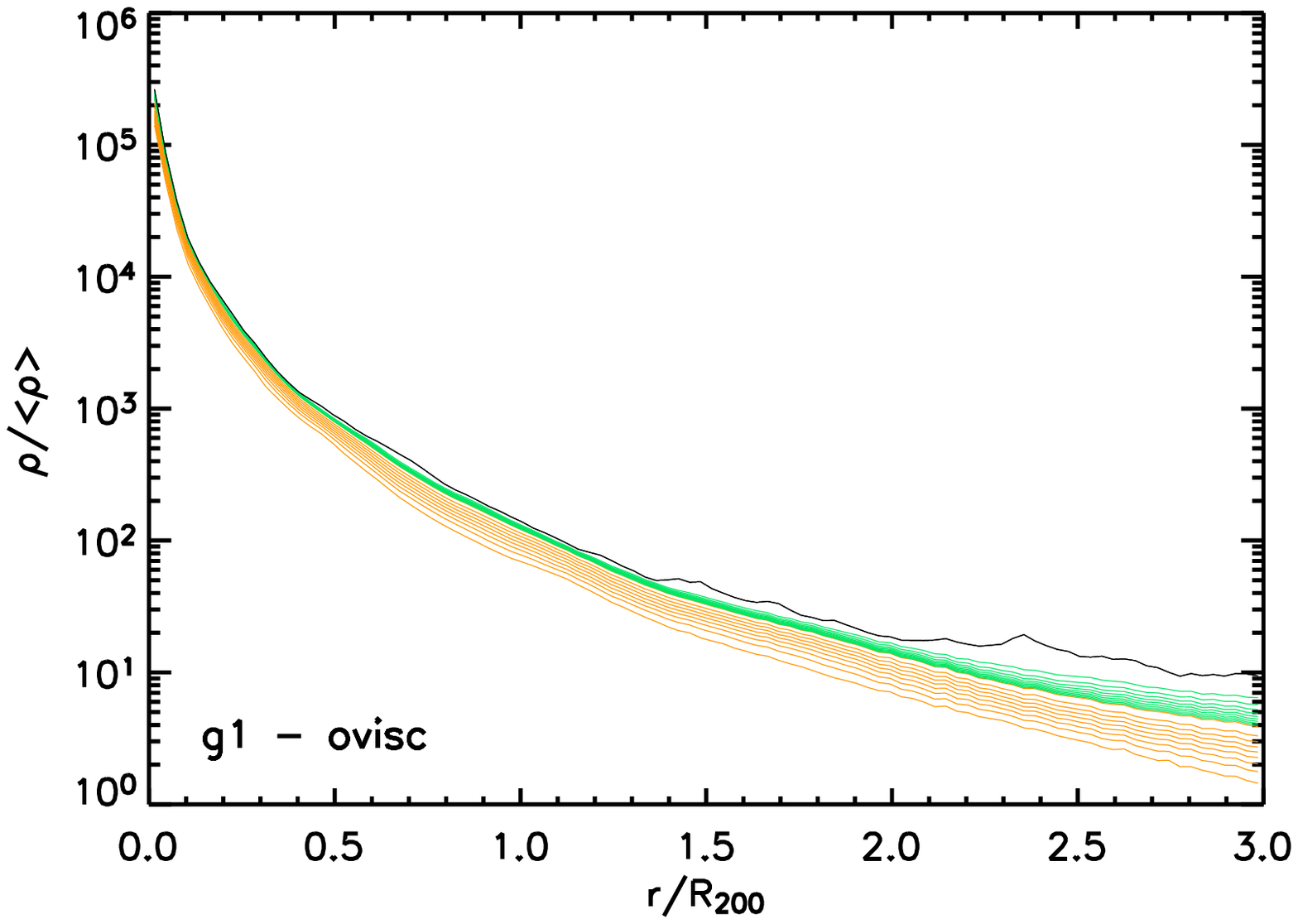}
\includegraphics[width=0.43\textwidth]{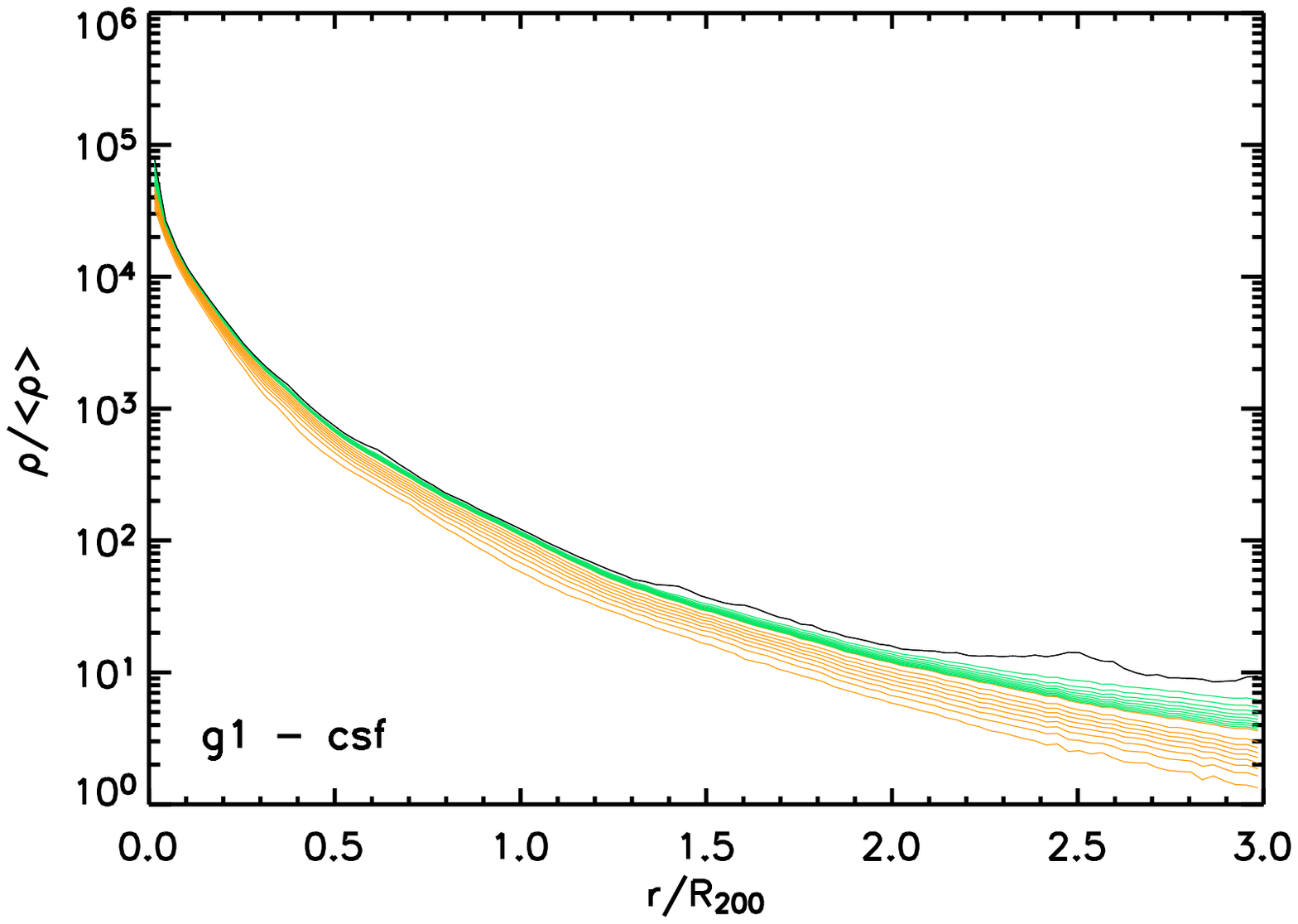}
\includegraphics[width=0.43\textwidth]{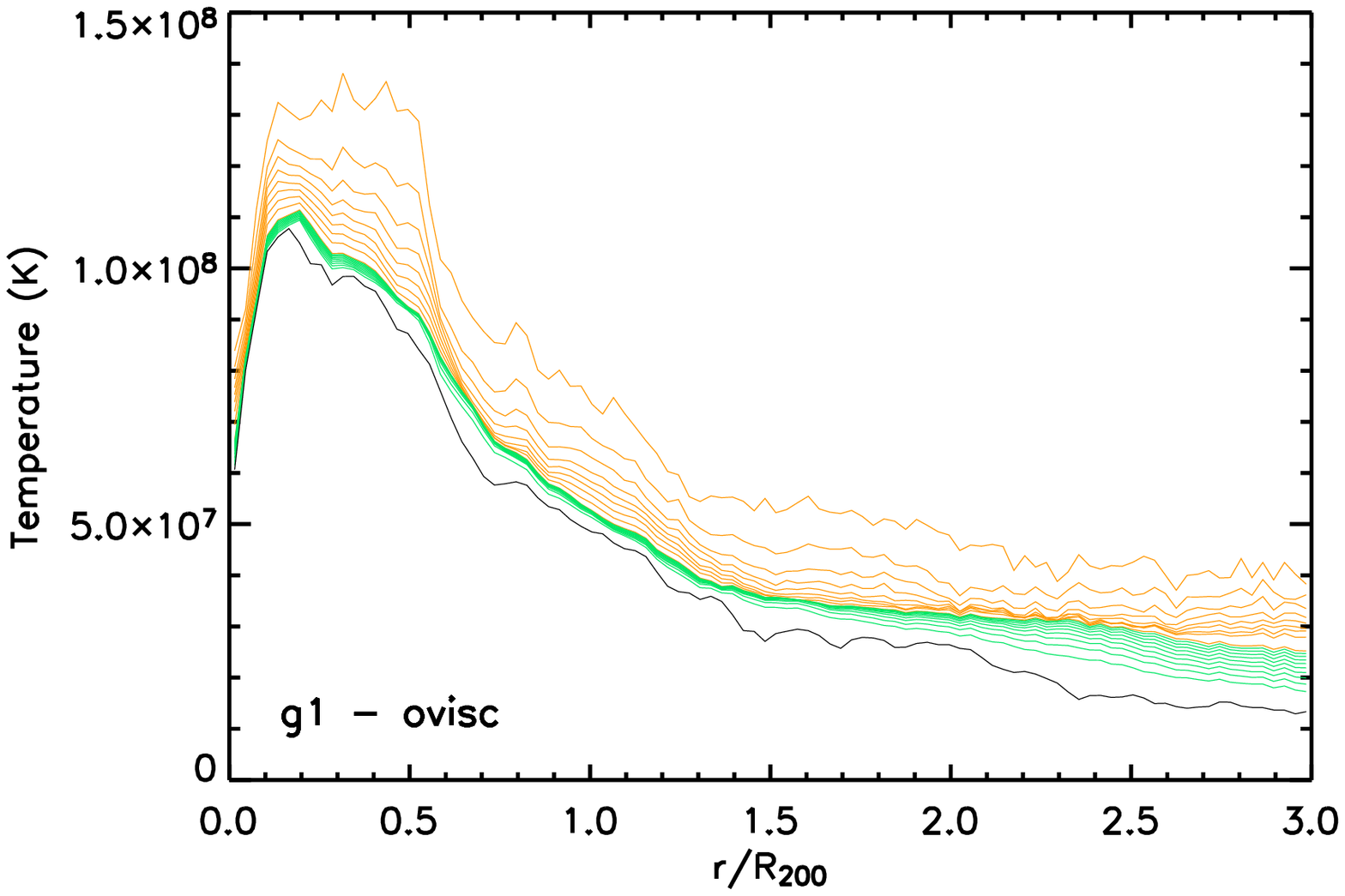}
\includegraphics[width=0.43\textwidth]{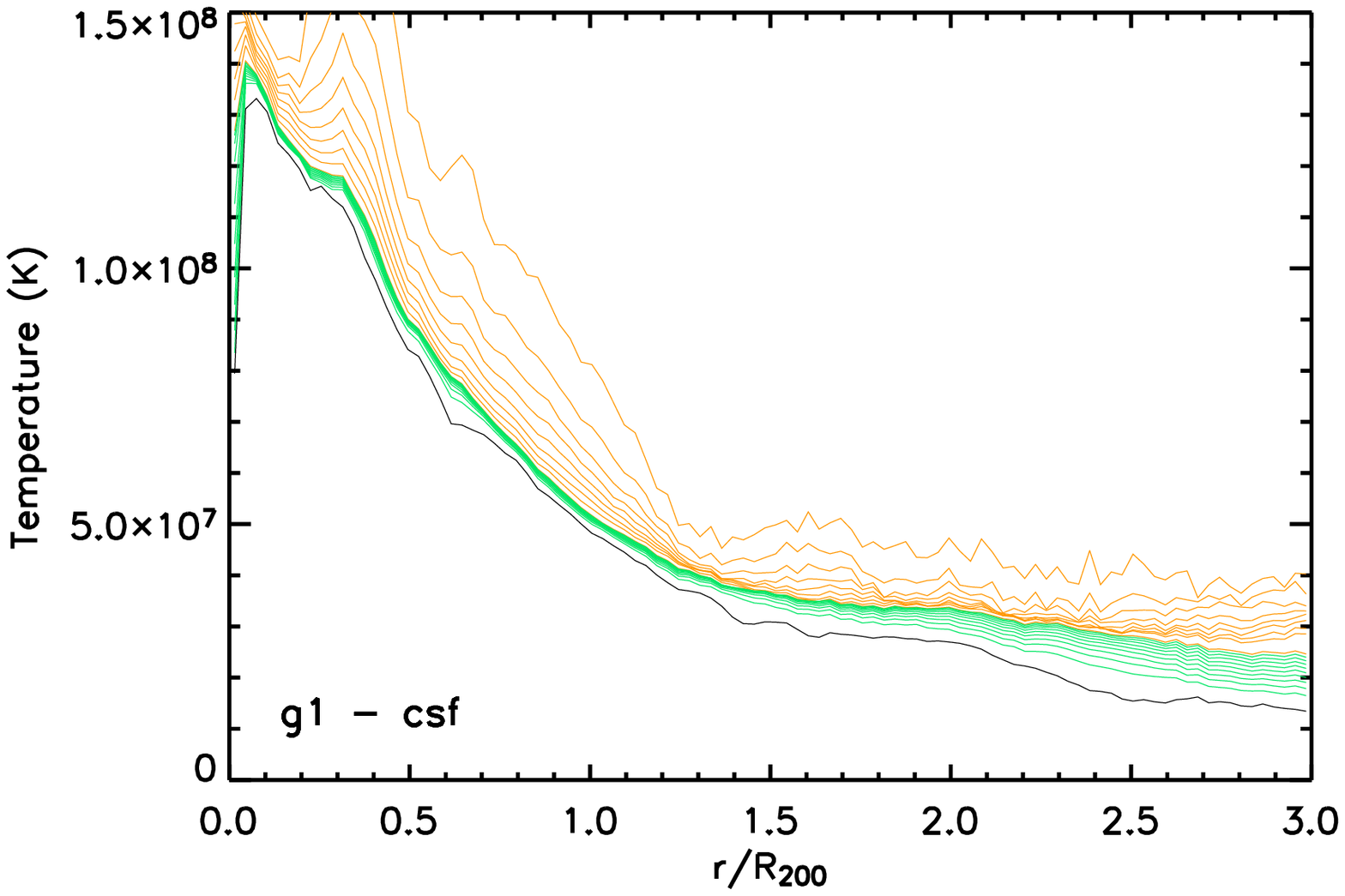}
\includegraphics[width=0.43\textwidth]{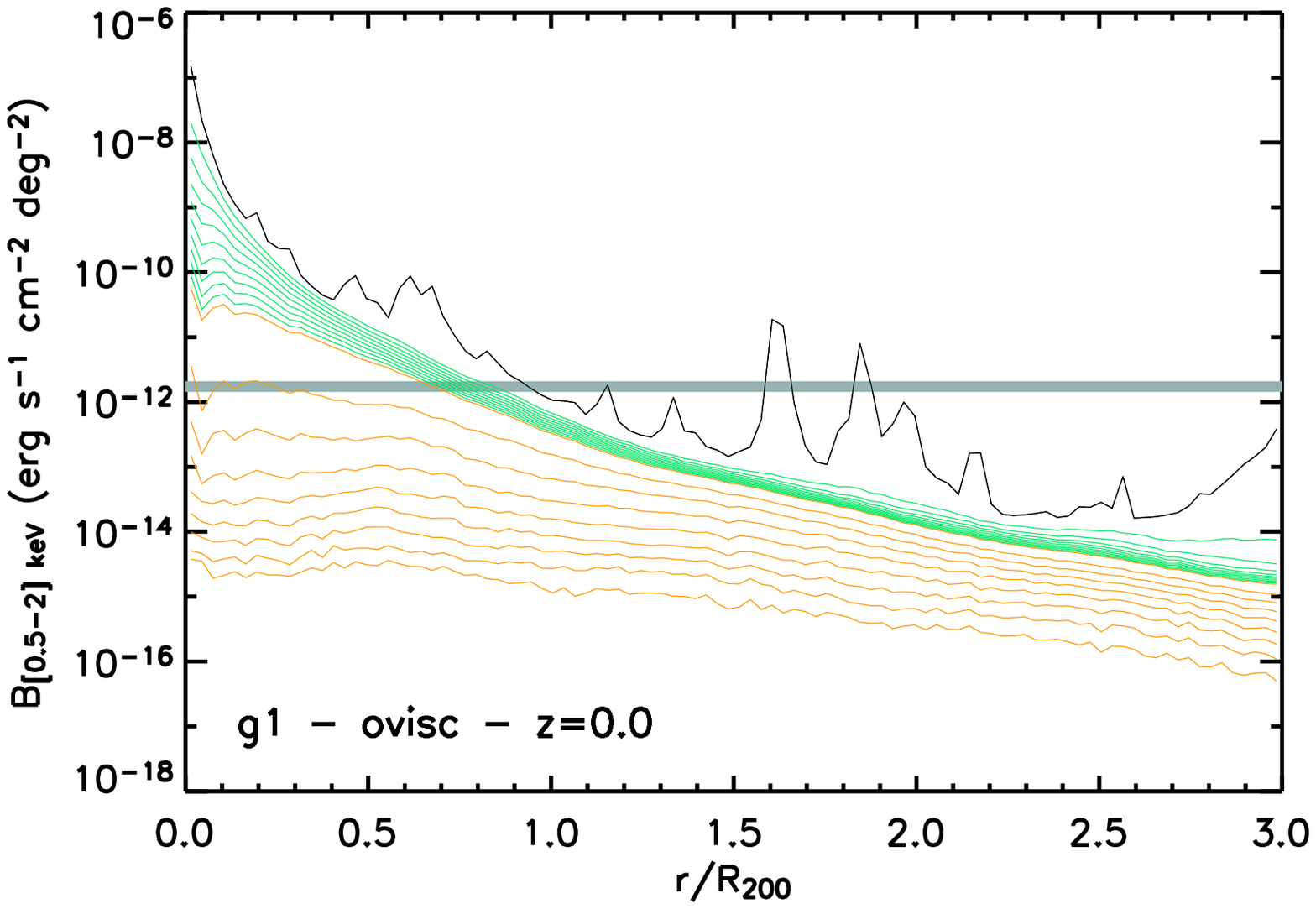}
\includegraphics[width=0.43\textwidth]{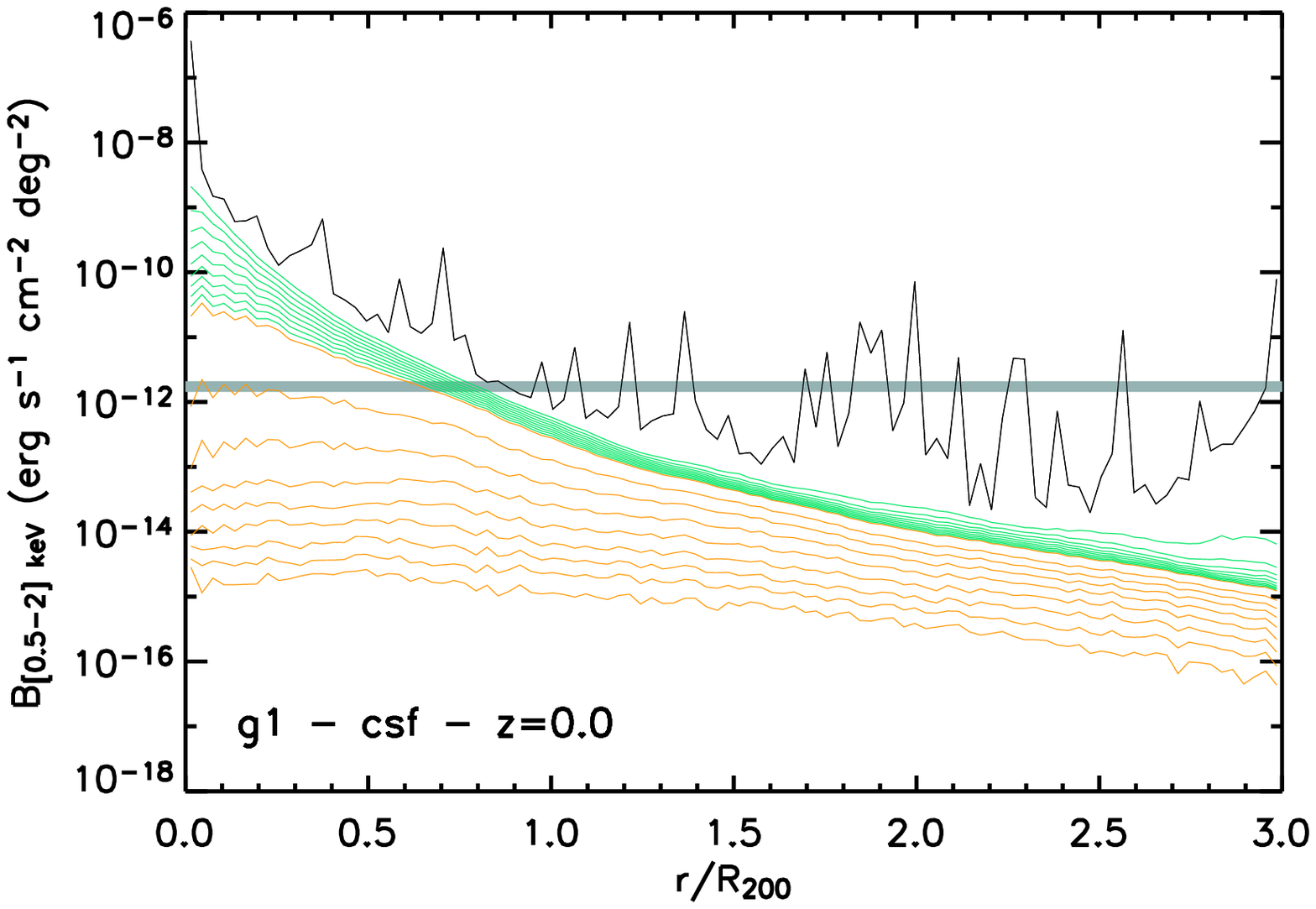}
\includegraphics[width=0.43\textwidth]{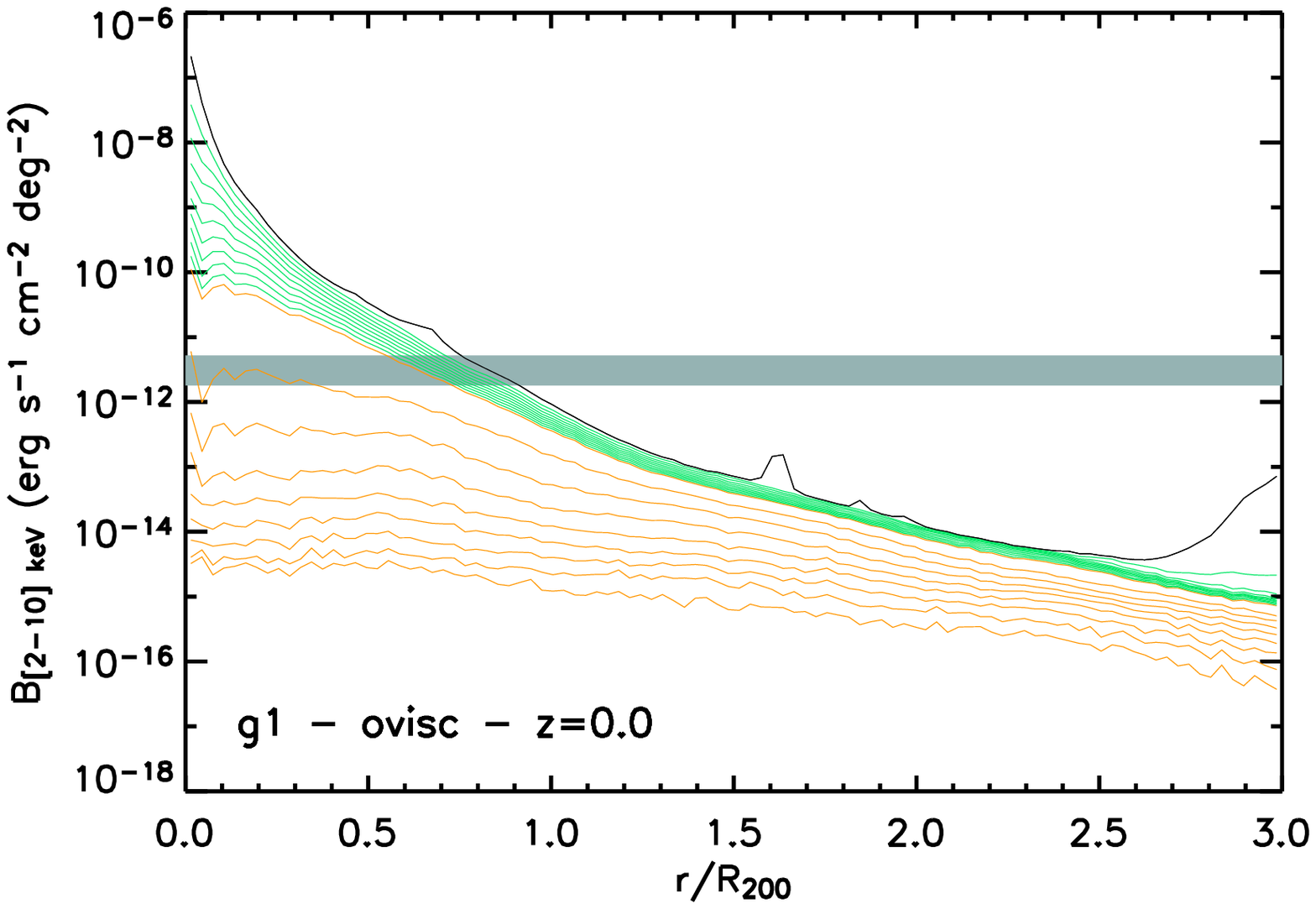}
\includegraphics[width=0.43\textwidth]{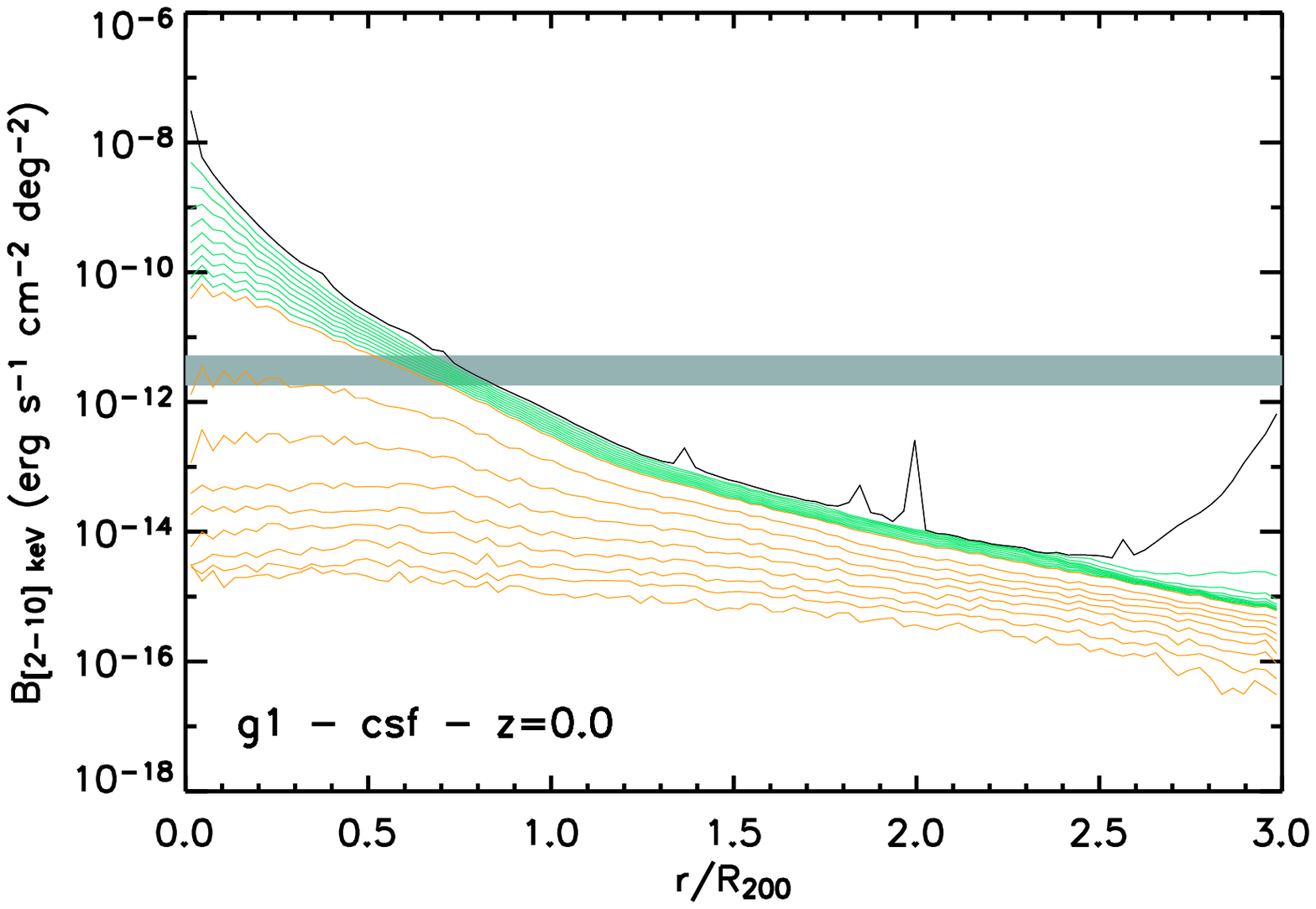}
\caption{
Radial profiles for different quantities as estimated for the {\it g1} cluster.
From top to bottom: gas density (normalized to the cosmic mean density, $<\rho> 
\equiv \Omega_{\rm b} \rho_{\rm c}$), mass-weighted temperature 
and X-ray surface brightness for the soft (0.5--2 keV) and the hard (2--10 keV) band 
calculated considering the clusters at $z=0$. 
Left column: \ovisc\ model. Right column: \csf\ model. These profiles are obtained from
different prescriptions: the black line shows the average of all the gas particles, 
the orange lines are averages of the different volume percentages from 10 to 90 per cent 
and the green lines are from 91 to 99 per cent. The extragalactic unresolved 
background from \protect \cite{hickox2006} in the [0.5--2] keV and [2--8] keV band of 
is indicated by the shaded region in the soft and hard X-ray surface brightness panels, 
respectively.
} \label{fig:profiles}
\end{figure*}

\subsection{Preparation of the dataset}

Focusing our attention on the cluster outskirts, we do not consider in our study
the effects of the cluster core on the overall properties, avoiding to deal 
with the theoretical, numerical and observational uncertainties on the physics of 
the X-ray emitting plasma in the cluster central regions.
Furthermore we must take into account that the mass enclosed in a given radial 
bin in the external regions of clusters ($r > R_{200}$) is dominated by the presence 
of dense subclumps of the typical dimension of galaxies or small galaxy-groups ($M 
\lesssim 10^{13} M_\odot$). Evidences of the existence of these clumps 
in the regions inside $R_{200}$ have also been shown by X-ray observations, but 
it is still not clear whether the frequency and the emission properties of the 
simulated clumps agree with the observed ones: this is due both to the low surface 
brightness of these objects that makes them difficult to resolve and to uncertainties 
in the theoretical modeling (e.g. simulations including cooling tend to produce more 
clumps and simulations with feedback from SNe end up with smaller clumps than 
non-radiative ones).

Anyway, since  
(i) we are interested only in the average properties of galaxy clusters and 
(ii) these small clumps are generally masked out in observational 
works, a direct comparison of our simulations with theoretical and observational results 
requires the introduction of a method to exclude these cold and dense clumps from 
our analysis.
It is quite difficult to define a precise and unique criterium 
to identify these condensed regions in the simulation volume from their dynamical and 
thermodynamical properties, therefore we decided to identify the volumes corresponding to 
non-collapsed regions. For this purpose we proceed in the following way: for every shell that 
corresponds to a radial bin of the profile, we sort the gas particles in decreasing volume order 
(i.e. starting from the more diffuse particles and going up to the denser ones, being 
the volume associated to the $i$-th particle defined as $V_i=m_{\rm gas}/\rho_{{\rm gas,}i}$)
then we compute the profiles calculating the mean of a given quantity considering only 
the particles summing up to a given percentage of the total volume of the shell.
In order to identify the physical properties of the regions excluded by our analysis,
we plot in Fig.~\ref{fig:maps} as an example the maps of the projected density, 
mass-weighted temperature and soft (0.5-2 keV) X-ray surface brightness
of the {\it g1} cluster considering (i) all the particles and 
(ii) those that are within the 95 per cent volume of each radial shell.
We can clearly see from these maps that the volume-selective method actually separates
the cold and dense clumps from the rest of the object. Anyway these maps show that this
criterion is much more effective in excluding dense regions than cold ones. However, 
since the X-ray emission is strongly dependent on the density of the gas, 
we can safely say that it well mimics the observational technique of masking the 
bright isolated regions. \\

We also study the shapes of X-ray surface brightness profiles of our simulated clusters. 
Since we are interested in the outskirts of galaxy clusters we cannot adopt a pure 
bremsstrahlung emission model because in the corresponding temperature ranges ($T<$ 1 keV) 
line emission gives a significant contribution. 
Therefore in order to calculate the profiles we adopt a MEKAL emission model 
\citep{mewe1985, liedahl1995}  assuming a gas metallicity $Z=0.2 Z_\odot$ 
\citep[using the solar value tabulated in][]{anders1989}
constant for every particle of the simulation: 
although this value can be an underestimate of the metallicity of the clusters' centre 
\citep[see][for more  detail]{degrandi2004}, the chosen model can be considered a good 
representation of the spectral properties 
of the clusters' external regions. In the following analyses we consider the 
objects at redshift $z=0$. To compute the surface brightness profiles, we adopted 
the same volume-selection scheme of the density and temperature profiles with the only 
difference that, since we want to obtain the surface brightness profiles of our objects,
we need to take into account the two-dimensional projection on the sky plane. 
To this purpose, we consider circular ring sections instead of spherical shells:
for every ring, we select all the particles that fill a given volume fraction $f$, we 
calculate their luminosity and we obtain $L_f$ by summing all of them. Finally we normalize 
it to take into account also for the remaining volume and use 
the normalized luminosity $L_{ring}=\frac{L_f}{f}$ to calculate the surface brightness 
(see the effects of this method on the [0.5-2] keV X-ray surface brightness 
shown in the bottom panel of Fig.~\ref{fig:maps}).
This method works efficiently for subtracting almost all the relevant 
substructures in the 
surface brightness profiles, except for two massive subhaloes present in the {\it g1} 
cluster simulation at a distance of 2-3 $R_{200}$ from the center; only for this cluster 
we excluded two separate angular regions (45 degrees wide each) from the computation of 
its X-ray surface brightness profiles.

In Fig.~\ref{fig:profiles}, we show the gas density, temperature and
surface brightness profiles into two different bands for a massive cluster ({\it g1}) 
assuming two different physical models: the ``ordinary-viscosity non-radiative"
(\ovisc) and the ``cooling + star formation + feedback" (\csf) one.
We plot the average profiles together with the ones obtained with different
cuts in volume.
The profiles in Fig.~\ref{fig:profiles} show that
eliminating the fraction of mass concentrated in 1 per cent of the volume
is sufficient to make the gas density, temperature and surface brightness
profiles much more regular than the average ones that consider all the particles
in the shell. For these reasons in the following analyses, we will mainly concentrate in 
fitting the 99 per cent--volume profiles and discuss the results obtained by cutting 
the volume at different percentages in Appendix~\ref{app:volume_cut}. By the way we also 
note that the temperature and density profiles obtained with this method for the \ovisc\ 
model well reproduce the shape of the theoretical profiles obtained 
by \cite{ascasibar2006}, if we assume a polytropic index 
$\gamma \simeq 1.18-1.20$.

\begin{figure*}
\includegraphics[width=0.43\textwidth]{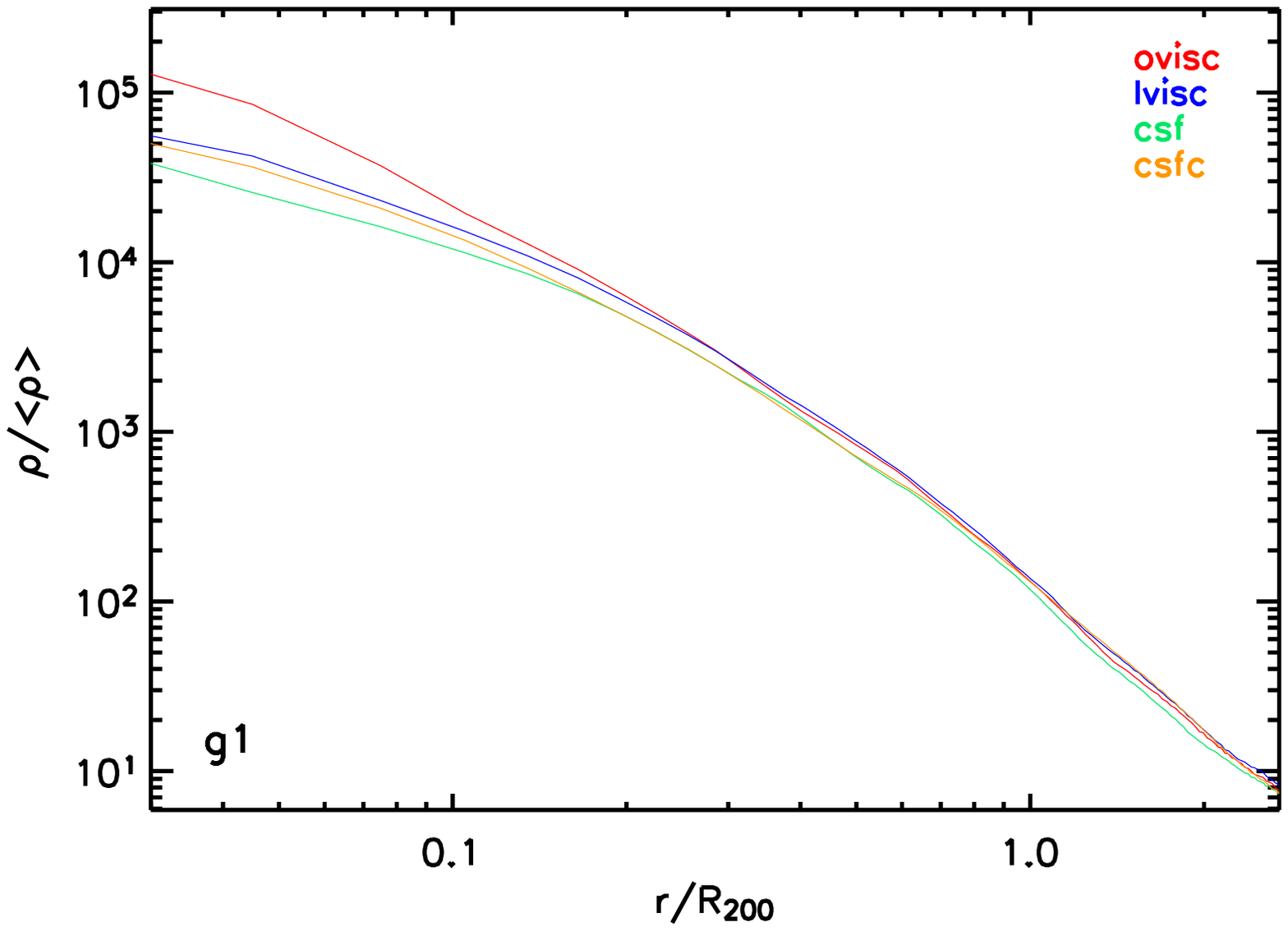}
\includegraphics[width=0.43\textwidth]{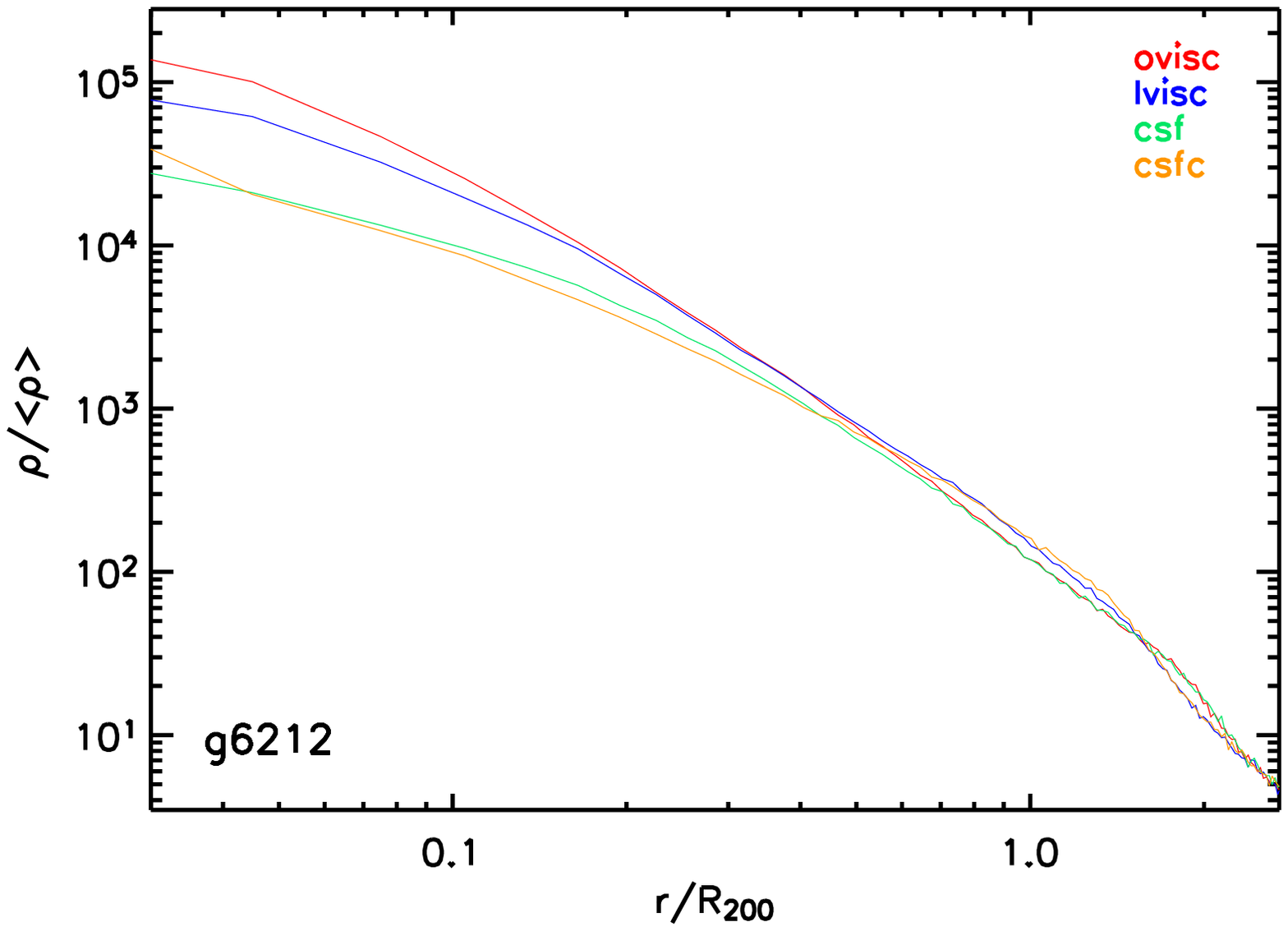}
\includegraphics[width=0.43\textwidth]{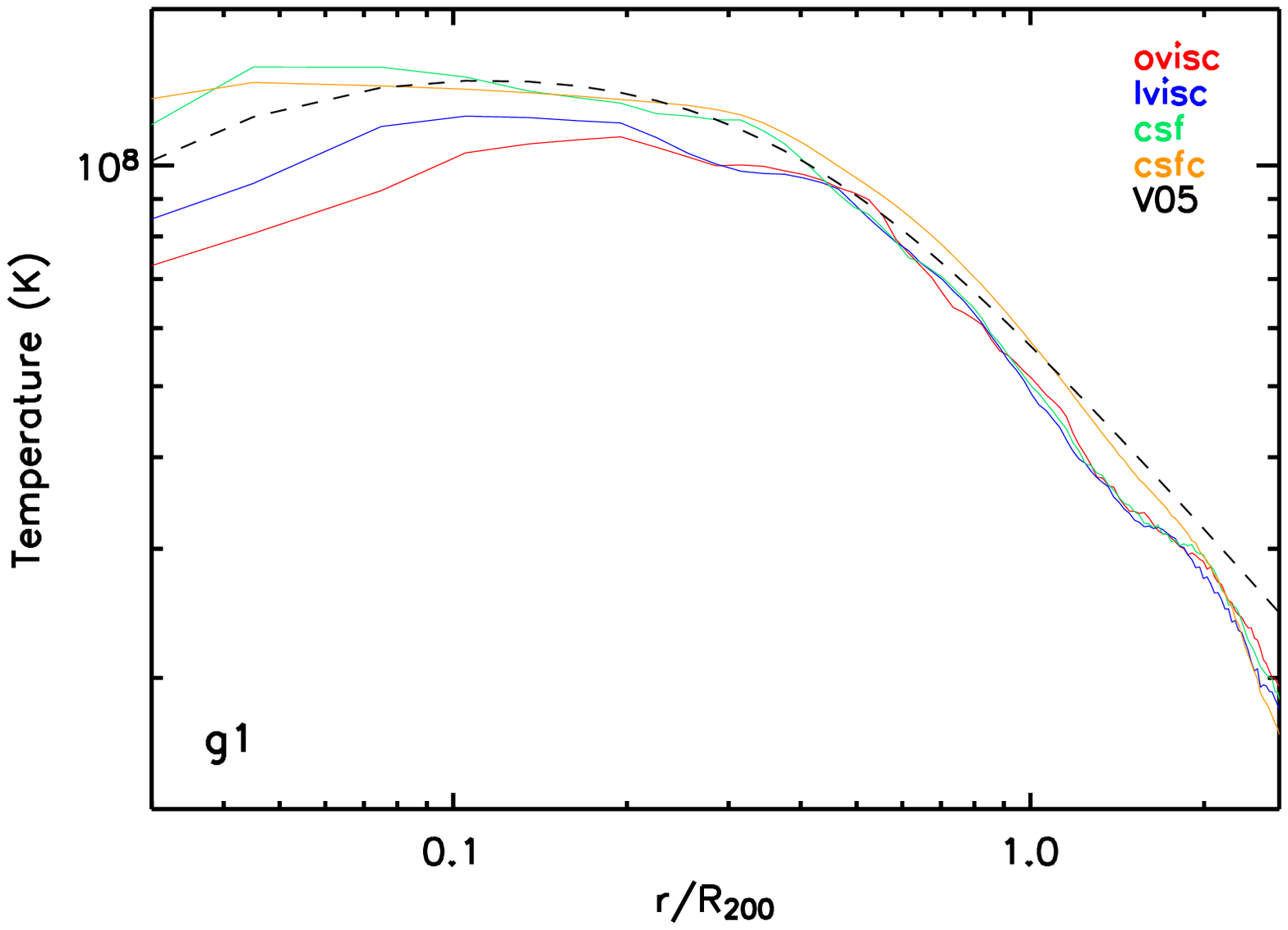}
\includegraphics[width=0.43\textwidth]{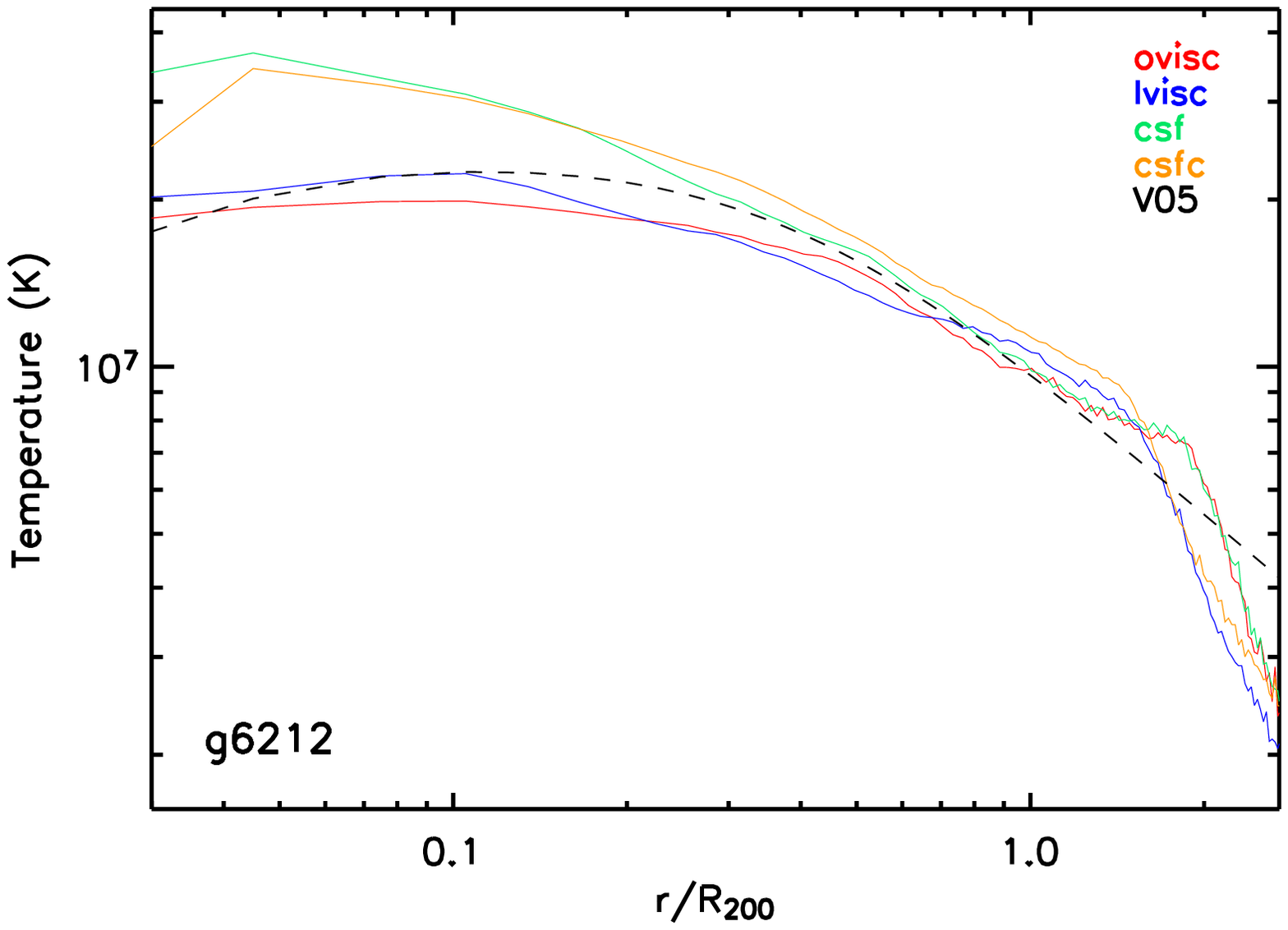}
\includegraphics[width=0.43\textwidth]{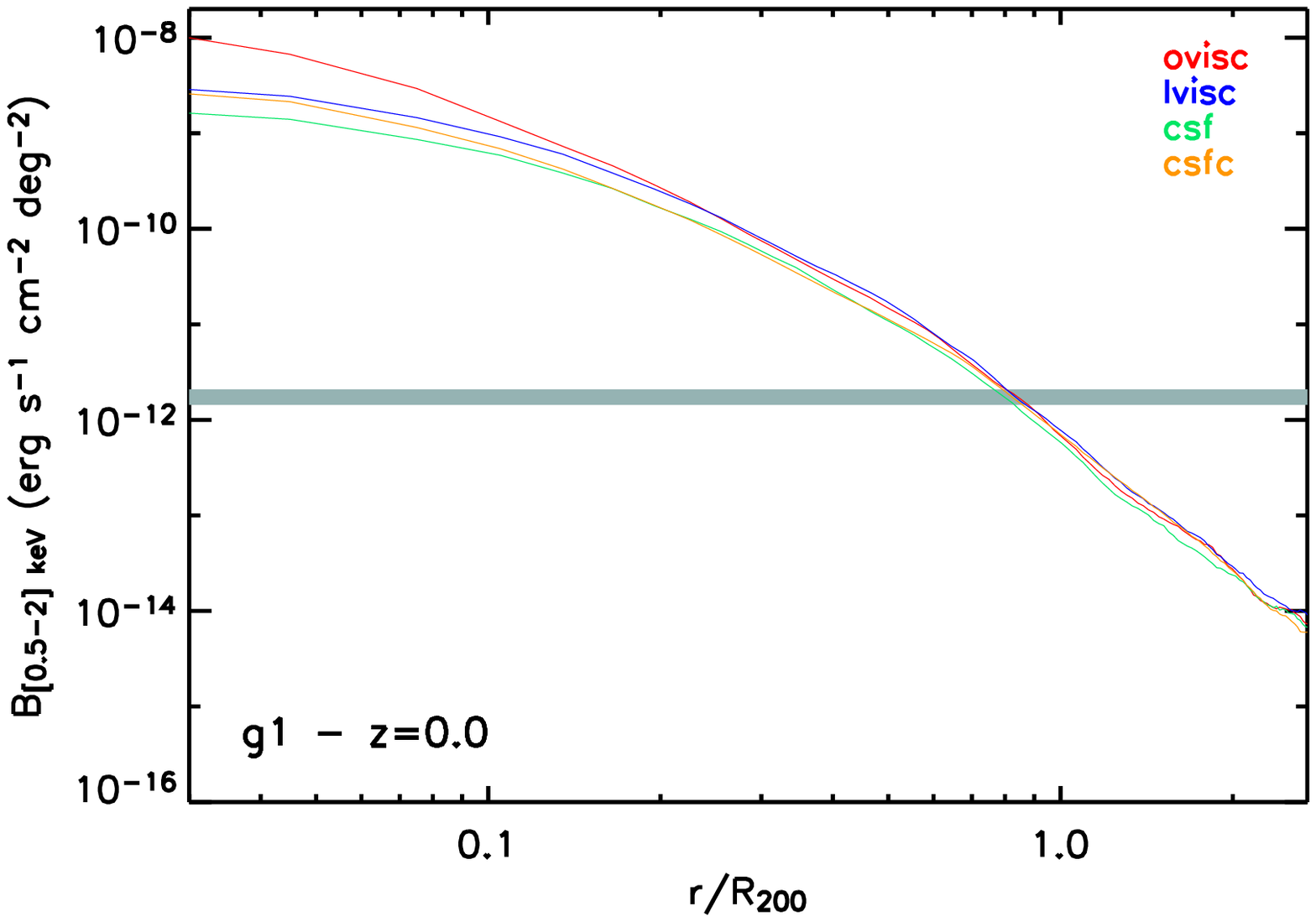}
\includegraphics[width=0.43\textwidth]{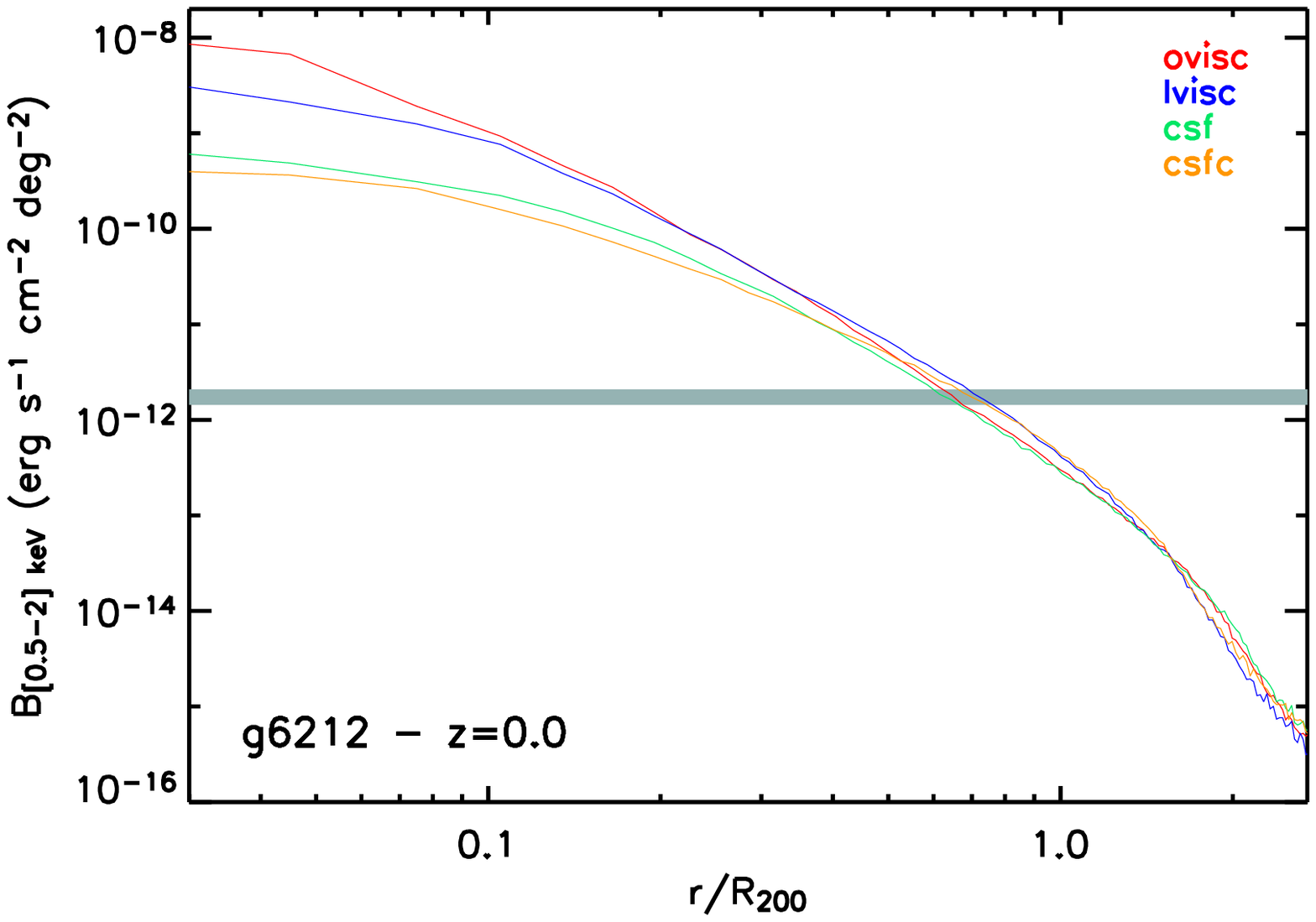}
\includegraphics[width=0.43\textwidth]{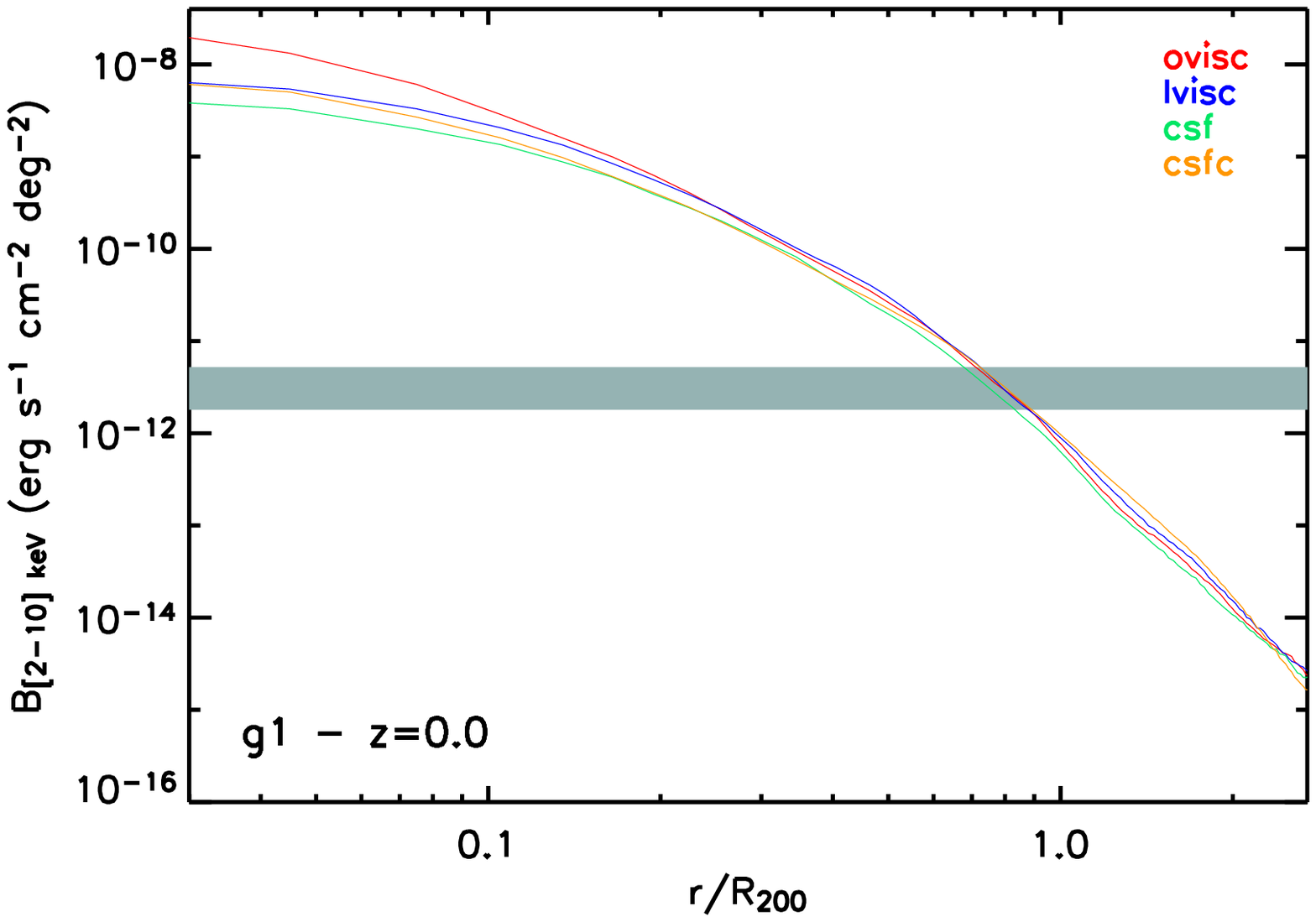}
\includegraphics[width=0.43\textwidth]{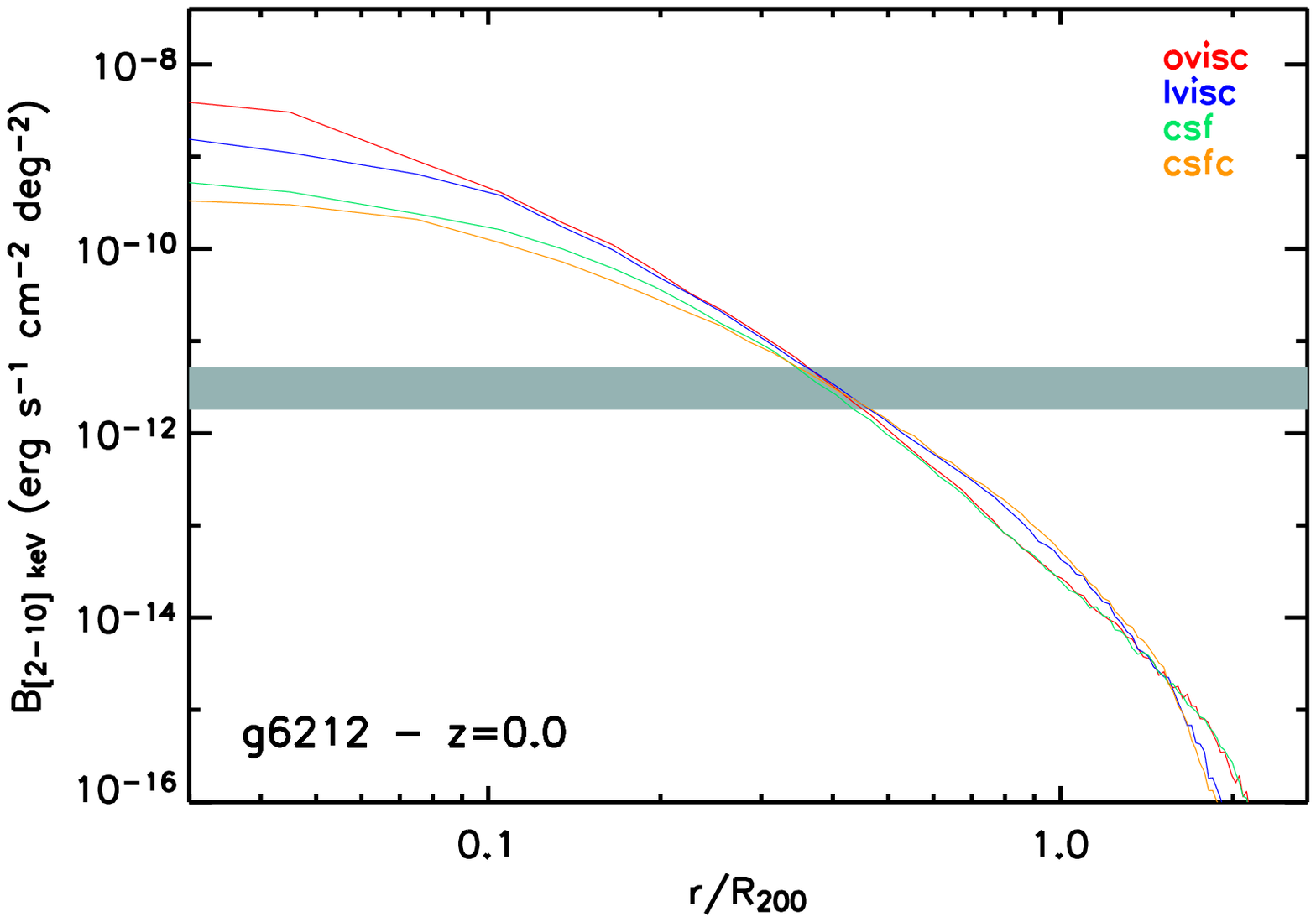}
\caption{
From top to bottom: comparison between the gas density (normalized to 
the cosmic mean density, $<\rho> \equiv \Omega_{\rm b} \rho_{\rm c}$), 
mass-weighted temperature, soft and hard X-ray surface brightness 
profiles extracted from the 99 per cent volume of the clusters {\it g1} 
(left column) and {\it g6212} (right column) simulated by using 4 different 
physical models. A dashed line indicates the functional form from 
\protect \citet[eq.~9]{vikhlinin2006} that well reproduces the 
behaviour of the temperature profile of nearby bright galaxy clusters 
observed with {\it Chandra}. The functional form is normalized to the 
average values of the temperature obtained from the 4 models at 
$0.5 R_{200}$. The extragalactic unresolved 
background from \protect \cite{hickox2006} in the [0.5--2] keV and [2--8] keV band of 
is indicated by the shaded region in the soft and hard X-ray surface brightness panels, 
respectively.
}
\label{fig:compare}
\end{figure*}

In Fig. \ref{fig:compare} we compare the 99 per cent volume profiles
for one of the largest and the smallest cluster of the samples using the four 
different physical models. As already noted in \cite{borgani2004}, 
radiative cooling and star formation selectively remove low-entropy gas from 
high-density regions. Consequently models with cooling tend 
to produce clusters less dense at the centre ($r<0.1 R_{200}$): in 
these regions we find that the gas density 
in non-radiative models is higher by a factor of $\sim 3.5$ in the 
most massive clusters and  more than 5 in less massive ones.
The same cooling processes generate a lack of pressure near the 
centre of the clusters that heats the infalling gas up to a factor of 
$\sim 1.8$ higher than in non-radiative simulations for all mass scales 
\citep[see also][]{tornatore2003}.
As a result, the X-ray emission in the internal regions is significantly 
reduced in the \csf\ and \csfc\ models. The external regions of the 
clusters are less affected by the different physics because the 
low-density gas is less prone to cooling and feedback effects.
As we will show later, this makes the results on the external regions 
of clusters only slightly dependent on the physical model adopted 
in the simulation \citep[see also][for influences of other physical 
processes]{romeo2006}.

\begin{figure*}
\includegraphics[width=0.43\textwidth]{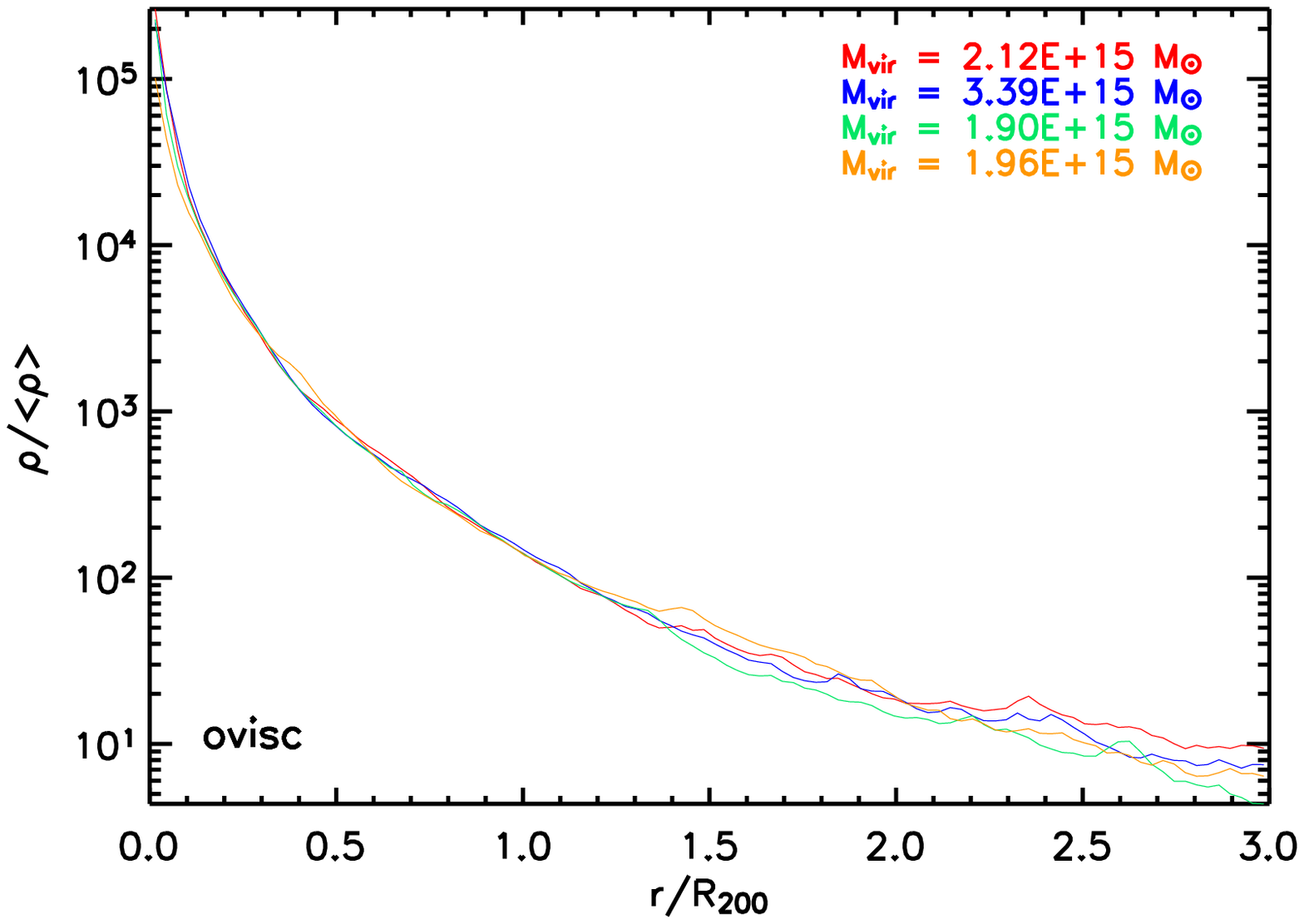}
\includegraphics[width=0.43\textwidth]{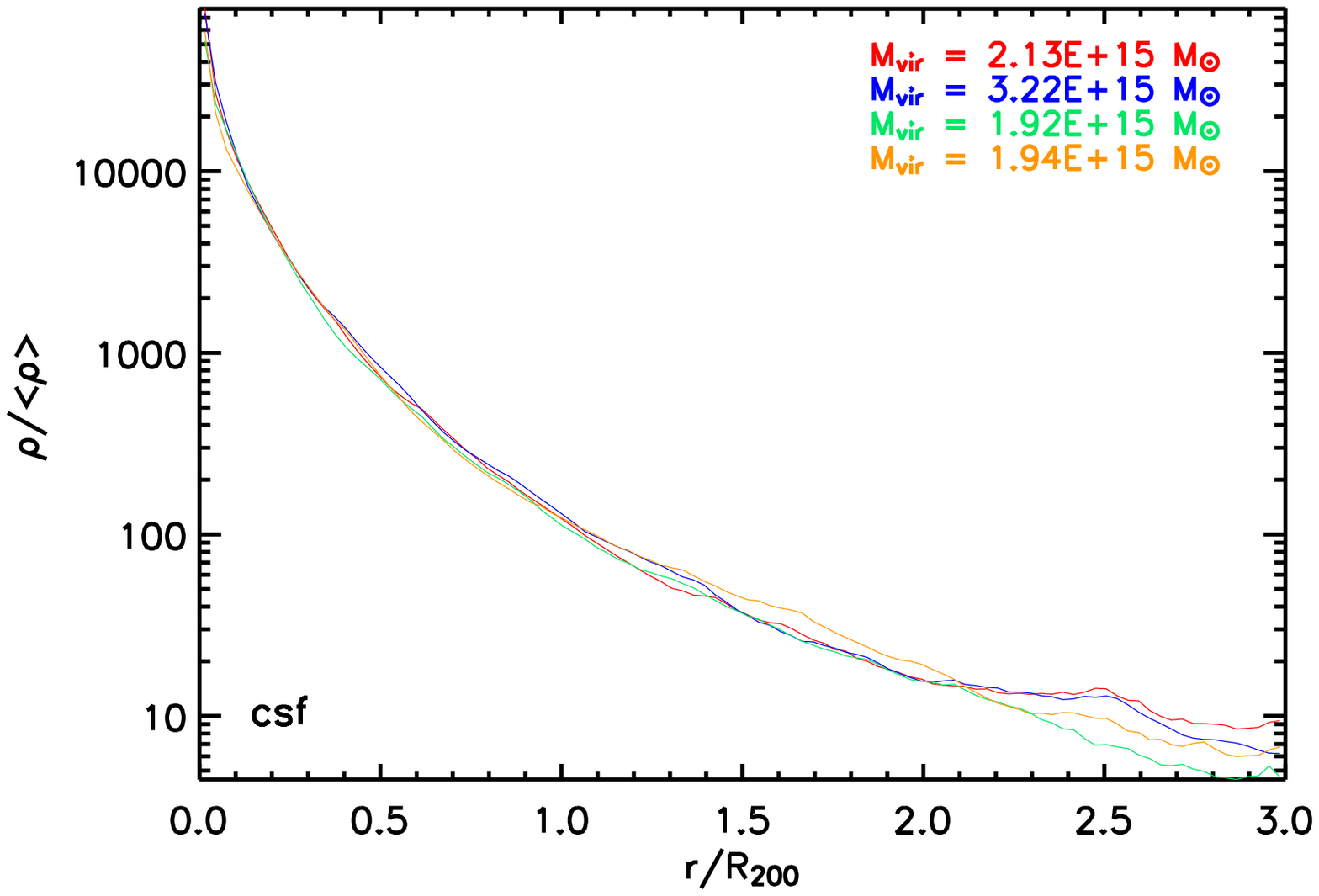}
\includegraphics[width=0.43\textwidth]{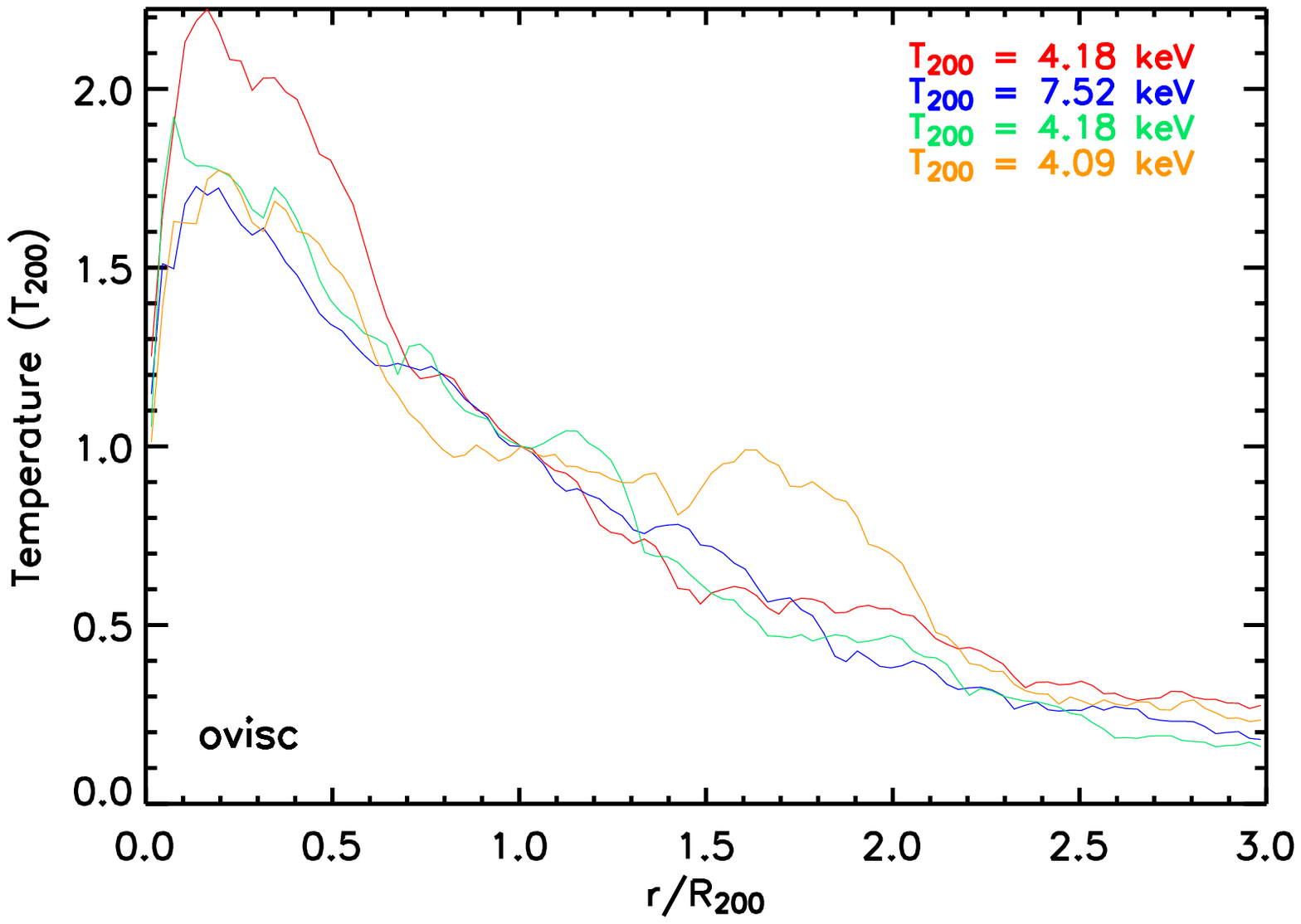}
\includegraphics[width=0.43\textwidth]{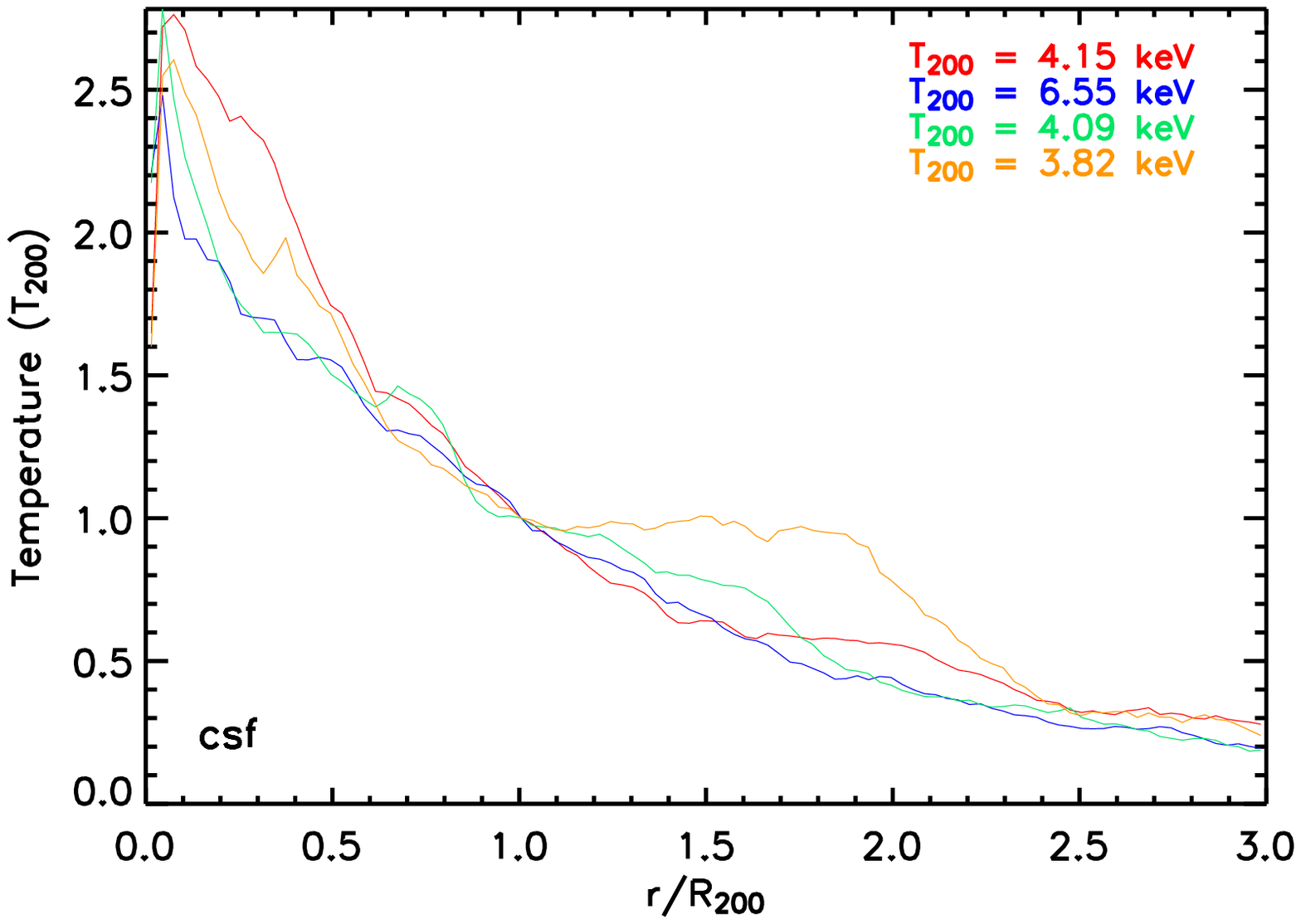}
\caption{Comparison between the 99 per cent volume profiles of density (normalized to 
the cosmic mean density, $<\rho> \equiv \Omega_{\rm b} \rho_{\rm c}$) (top panels) and 
mass-weighted temperature normalized to $T_{200}$ (lower panels) for the clusters {\it g1} (red), 
{\it g8} (blue), {\it g51} (green), {\it g72} (orange). Left and right columns refer to 
the \ovisc\ and \csf\ model, respectively.}
\label{fig:pr_compare}
\end{figure*}

In Fig.~\ref{fig:pr_compare} we compare the 99 per cent  volume profiles for the 4 most 
massive clusters of the dataset (sample $A$). The density profiles have a very 
similar shape up to about $2 R_{200}$ 
indicating a universal function. The regularity of the 99 per cent volume 
profiles extends also for distances $> 3 R_{200}$. The temperature profiles are more 
irregular (note that the temperatures are normalized using $T_{200}$, the temperature
at $R_{200}$ ): the bumps present in the {\it g1} profile 
near the centre and in {\it g72} at $r \sim 1.7 R_{200}$ are an 
effect of shocks at $r \sim 0.6 R_{200}$ and $r \sim 2.1 R_{200}$, 
respectively, due to recent major mergers.

In Table \ref{tab:prof_values} we show the ratios between the values of the 99 per cent volume profiles 
at different radial distances and their values at $0.3 R_{\rm vir}$ obtained with the \ovisc\ model. The 
regularity of these profiles can be seen by the low dispersion between the values of the clusters of 
the same sample.

\begin{table*}
\begin{center}
\caption{
Ratios between the profile values at different radial distances and their values at $0.3 R_{200}$. We show 
the average and the standard deviation between the two cluster samples for the \ovisc\ model.}
\begin{tabular}{lcccccc}
Radial distance ($R_{200}$)  &    0.3    &    0.5    &    0.7    &    1.0    &    2.0    &    3.0      \\
\hline

 & \multicolumn{6}{c}{Sample $A$}                               \\
$\rho$                  & $1.0$ & $0.31\pm0.02$ & $0.127\pm0.004$ & $(4.8\pm0.2)\times10^{-2}$ & $(5.6\pm0.8)\times 10^{-3}$ & $(1.7\pm0.5)\times 10^{-3}$  \\
$T$                     & $1.0$ & $0.88\pm0.06$ & $0.72\pm0.06$   & $0.60\pm0.06$              & $0.34\pm0.09$               & $0.16\pm0.03$                \\
$SB_{[0.5-2]{\rm keV}}$ & $1.0$ & $0.17\pm0.03$ & $(4.0\pm1.1)\times10^{-2}$ & $(8.4\pm1.2)\times10^{-3}$ & $(2.3\pm0.9)\times10^{-4}$ & $(6.3\pm4.3)\times 10^{-5}$  \\
$SB_{[2-10]{\rm keV}}$  & $1.0$ & $0.15\pm0.02$ & $(3.2\pm0.1)\times10^{-2}$ & $(5.5\pm0.1)\times10^{-3}$ & $(7.1\pm1.8)\times10^{-5}$ & $(6.4\pm5.7)\times 10^{-6}$  \\
\hline

 & \multicolumn{6}{c}{Sample $B$}                               \\
$\rho$                  & $1.0$ & $0.31\pm0.02$ & $0.13\pm0.02$ & $(4.8\pm0.7)\times10^{-2}$ & $(6.0\pm0.2)\times 10^{-3}$ & $(1.6\pm0.2)\times 10^{-3}$  \\
$T$                     & $1.0$ & $0.84\pm0.04$ & $0.72\pm0.04$   & $0.58\pm0.03$              & $0.29\pm0.06$               & $0.14\pm0.02$                \\
$SB_{[0.5-2]{\rm keV}}$ & $1.0$ & $0.16\pm0.02$ & $(4.2\pm1.1)\times10^{-2}$ & $(7.9\pm1.0)\times10^{-3}$ & $(1.8\pm0.9)\times10^{-4}$ & $(1.5\pm1.0)\times 10^{-5}$  \\
$SB_{[2-10]{\rm keV}}$  & $1.0$ & $0.11\pm0.01$ & $(2.1\pm0.7)\times10^{-2}$ & $(2.3\pm0.4)\times10^{-3}$ & $(1.7\pm1.4)\times10^{-5}$ & $(5.6\pm7.2)\times 10^{-7}$  \\
\hline

\end{tabular}
\label{tab:prof_values}
\end{center}
\end{table*}

\section{Results on the outer slopes of the radial profiles} \label{sect:results}

\begin{figure*}
\includegraphics[width=0.43\textwidth]{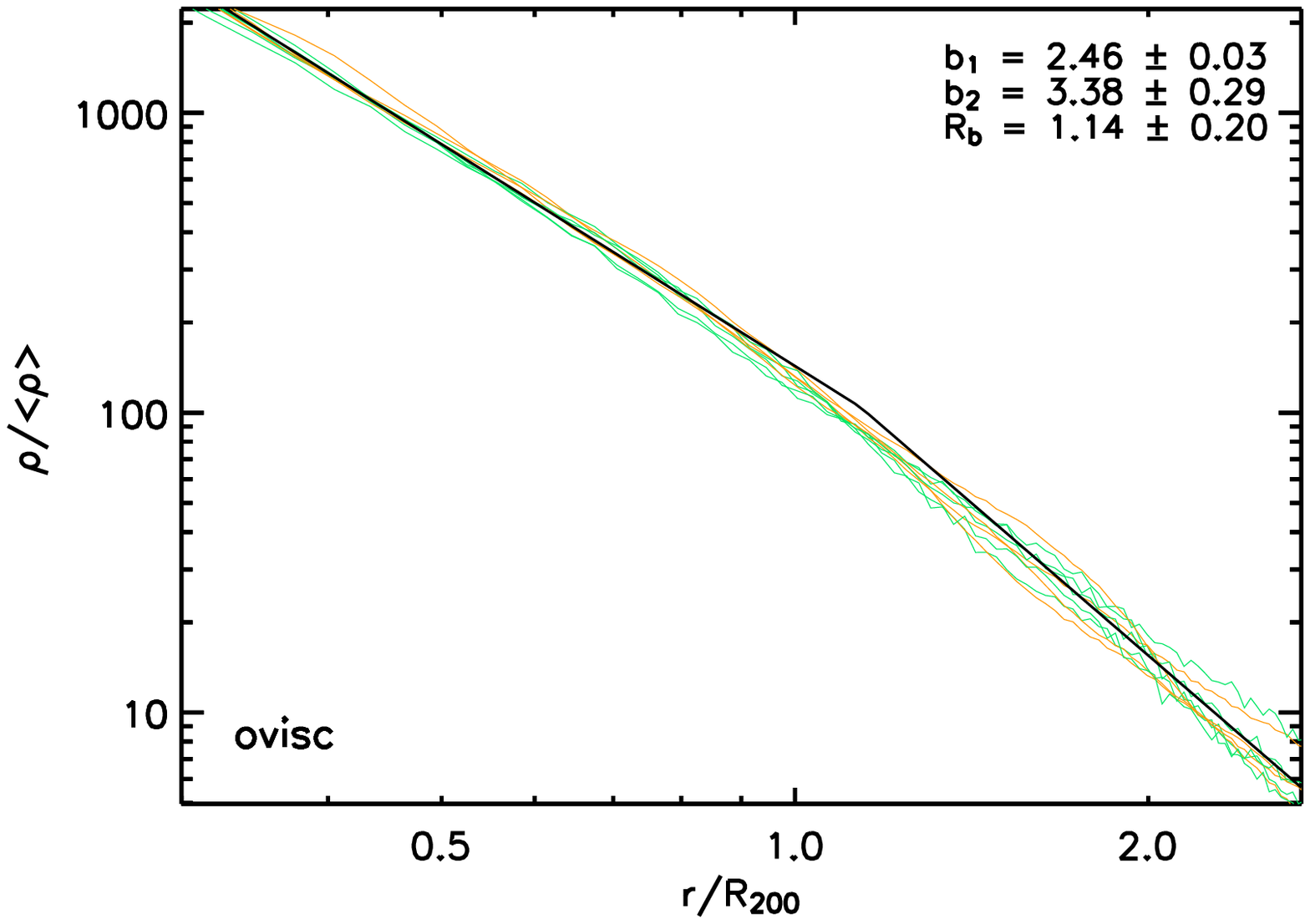}
\includegraphics[width=0.43\textwidth]{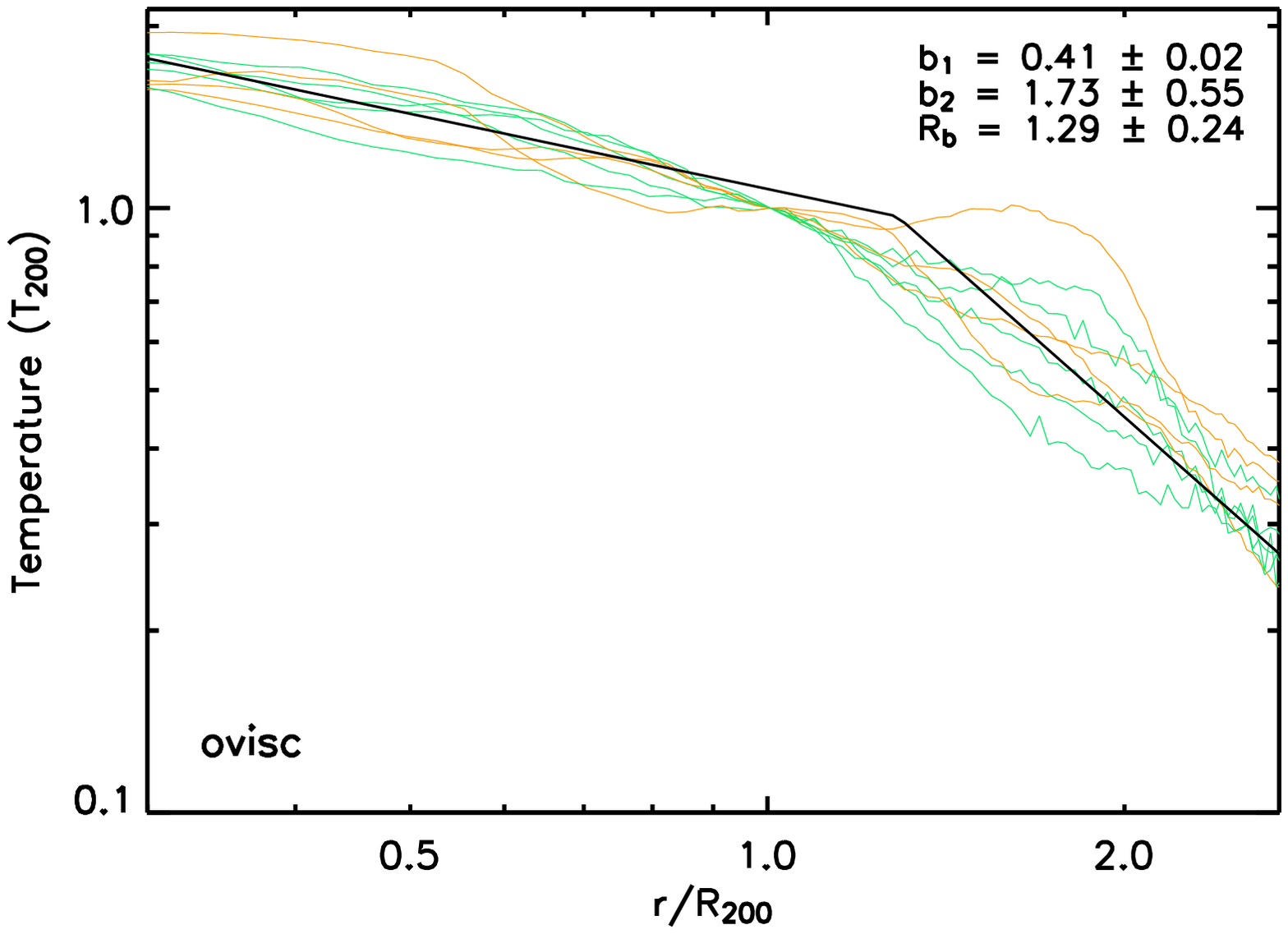}
\caption{Profiles (99 per cent volume) of the 9 clusters obtained with the \ovisc\ model.
The orange lines are for the 4 clusters of sample $A$, the green lines are for the 5 groups of 
sample $B$. We fit the data with a broken power-law relation of eq. \ref{eq:broken_pow_law}
in the interval $0.3 \leq r/R_{\rm 200} \leq 2.7$ and plot (black line) the relation assuming the 
average of the best-fit parameters between the 9 objects.}
\label{fig:pr_fit}
\end{figure*}

To describe the behaviour of the radial density and temperature profiles in the cluster outskirts,
we adopt a broken power-law relation given by the expression
\begin{equation}
y(x) = \left\{ \begin{array}{ll}
                  a x^{-b_1} & \mbox{if $x \le R_b/R_{200} $} \\
                  a \left(\frac{R_b}{R_{200}}\right)^{-(b_1-b_2)} x^{-b_2} & \mbox{if $x > R_b/R_{200} $}
                                 \end{array} \right. \ ,
\label{eq:broken_pow_law}
\end{equation}
in its \emph{logarithmic} form, where $r$ is the distance from the centre, 
$x \equiv r/R_{200}$ and $a$, $b_1$, $b_2$ and $R_b$ are the free 
parameters, representing the normalization, the inner slope, the outer slope and the 
radius at which the slope changes, respectively. 
The fit is computed in the interval $0.3 \leq x \leq 2.7$. 
We fit the profiles for the 9 clusters separately and then calculate the average and dispersion 
for every parameter. The best fit relations for the \ovisc\ model are shown together with the 
profiles of single clusters in Fig. \ref{fig:pr_fit}. 

The gas density profiles show a change in the slope of the power law 
at about 1.1-1.3 $R_{200}$, with an internal slope of 2.4-2.5 with no
significant dependence on the physical model or the volume selection scheme 
adopted (see the Appendix for details on the effect of different volume 
cuts) and also a very low dispersion between the different objects. This 
indicates that the behaviour of the density profiles is well determined 
from outside the core to more than $R_{200}$.
In the outskirts the slope increases up to $b_2\sim 3.4$ for all the models 
showing more dispersion between the 9 haloes due to the fact that in these regions 
the influence of the environment is becoming more important with respect to the 
potential well of the cluster itself.

In the temperature profiles, the steepening towards the external regions is more 
prominent and the cutoff is around $r=1.3-1.5 
R_{200}$ with significant dispersion between the clusters. The best fit internal 
slope is significantly affected by the physics: the profiles obtained 
with non-radiative simulations have slopes of $b_1\sim 0.4$ while \csf\ and \csfc\ models 
produce steeper profiles with slopes of $b_1\sim 0.6$. Also in the outskirts the non-radiative 
models tend to produce slightly shallower profiles with the difference less significant 
due to the higher dispersion between the different haloes. The different slopes 
of the profiles is an effect of the cooling that creates an higher gradient of temperature 
as can be seen in Fig. \ref{fig:compare}.

In Fig.~\ref{fig:profiles}, we also show the soft (0.5--2 keV) and hard (2--10 keV) X-ray
surface brightness profiles of the \emph{g1} object for the different physical models. 
These plots indicate that models with cooling produce a higher number of
clumps that dominate the soft X-ray emission in the external regions.
These clumps are indeed those that are well eliminated by our selection in volume.
We also obtain that the cluster surface brightness at $z=0$ reaches the value of the 
unresolved X-ray background estimated by \cite{hickox2006} of $1.76 \pm 0.32$ in the [0.5--2] 
keV band and $3.5 \pm 1.7$ in the [2--8] keV band in units of $10^{-12}$ \sbunits,
at approximately $0.8 R_{200}$ and $0.4 R_{200}$ in the soft and hard band, respectively; 
considering the effect of redshift dimming and that, according to the 
recent results of \cite{roncarelli2006}, a significant 
fraction of this background can be due to the diffuse gas of non-virialized objects, the 
perspective of observing the regions around the clusters virial radius is extremely 
challenging.

\begin{figure*}
\includegraphics[width=0.43\textwidth]{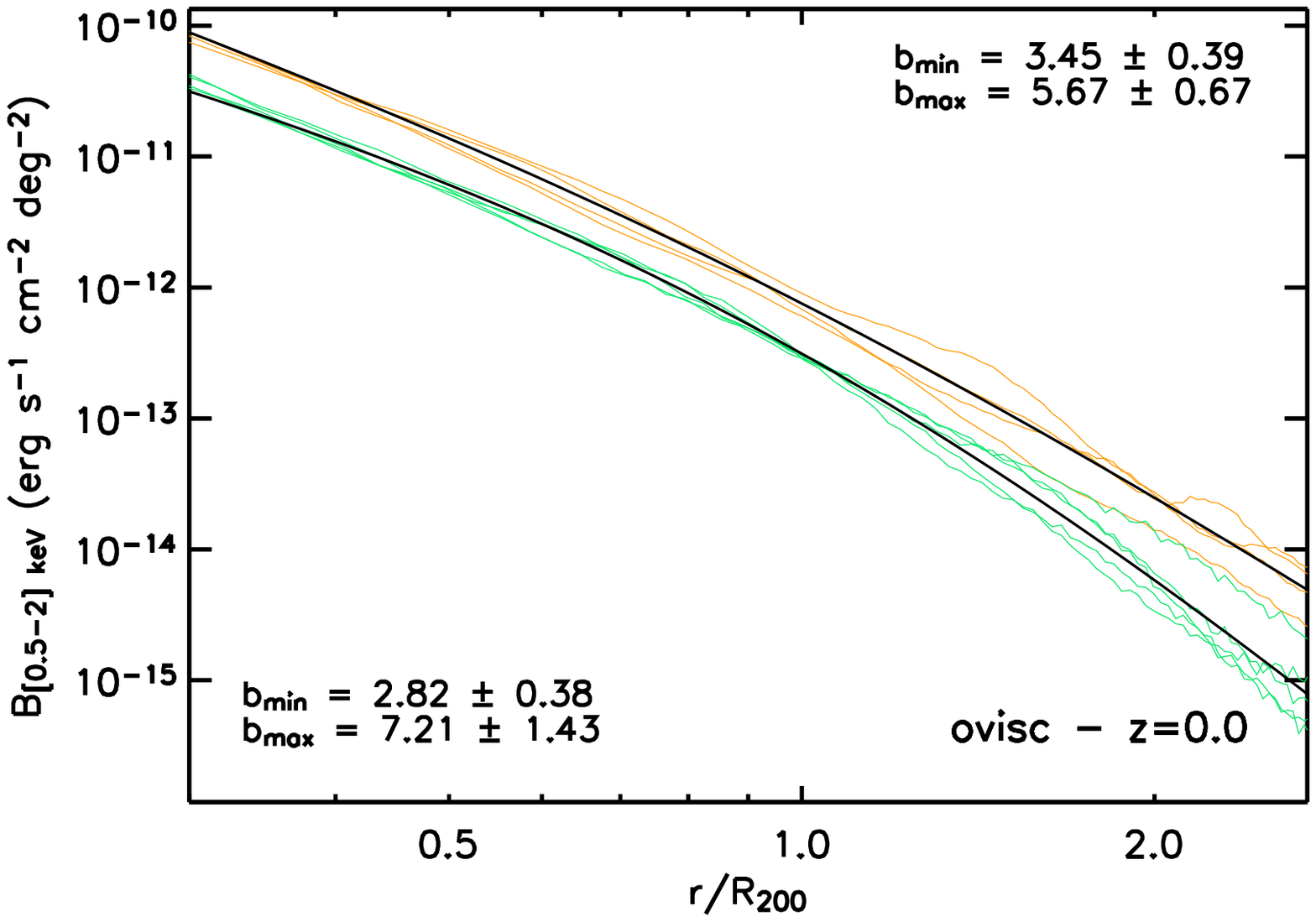}
\includegraphics[width=0.43\textwidth]{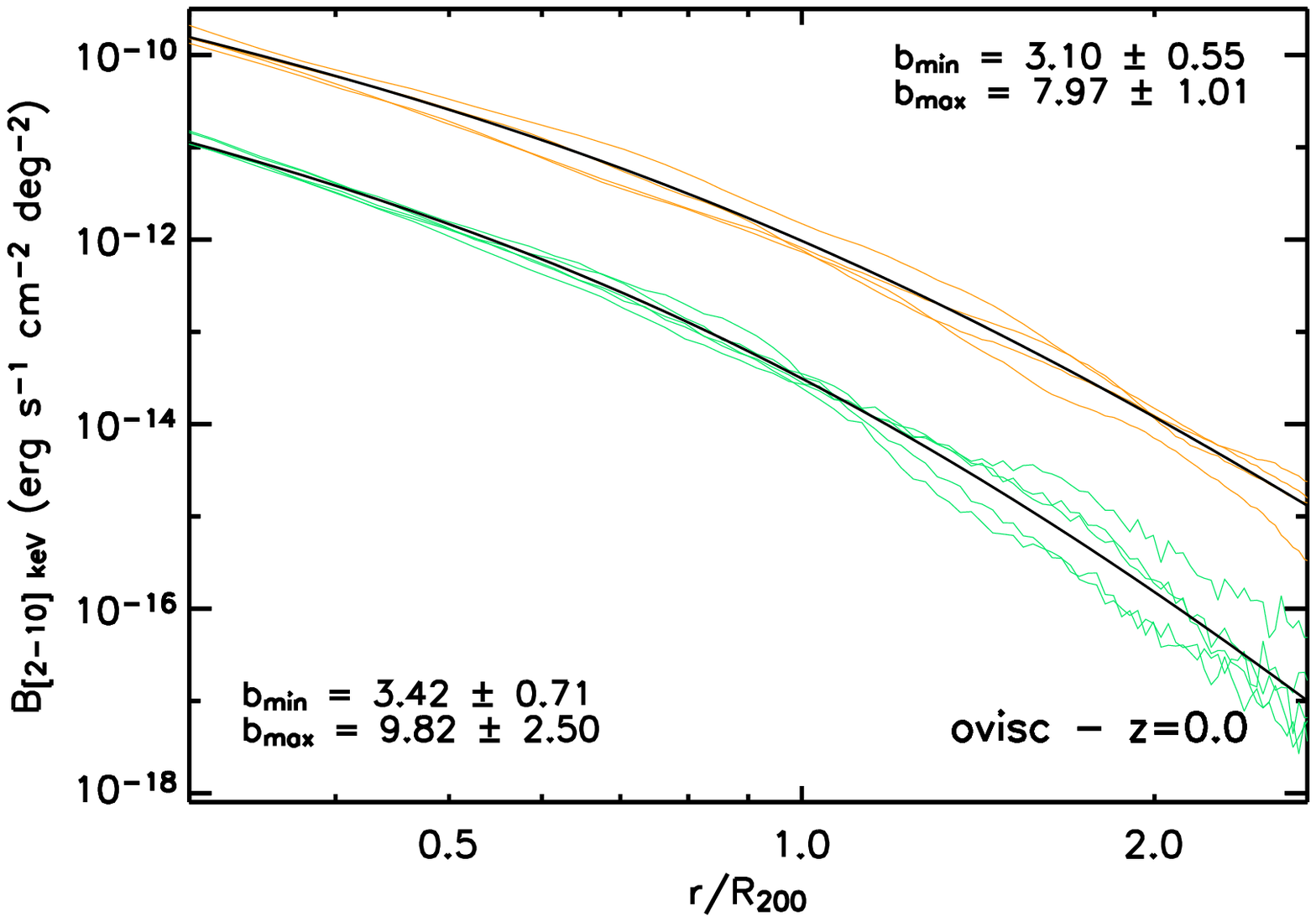}
\caption{X-ray surface brightness profiles (99 per cent volume, [0.5--2] and [2--10] keV band) 
of the 9 clusters for the \ovisc\ model. The orange lines are for the 4 clusters of sample $A$, 
the green lines are for the 5 groups of sample $B$.
We fit the data of the two samples separately with a rolling-index power-law relation of 
eq. \ref{eq:roll_pow_law} in the interval $0.3 \leq r/R_{\rm 200} \leq 2.7$ and plot (black lines) 
the relations assuming the average of the best-fit parameters between the objects of the samples.}
\label{fig:fit_rpl}
\end{figure*}

\begin{table*}
\begin{center}
\caption{
Distribution of the best-fit values of the rolling-index power-law functions (see eq.
\ref{eq:roll_pow_law}) of the soft (0.5--2 keV) and hard (2--10 keV) X-ray surface brightness
profiles separated between the two samples $A$ and $B$ and for different physical models. 
The profiles are extracted from the 99 per cent volume-selection scheme and fitted over the radial 
range $0.3 \leq r/R_{\rm 200} \leq 2.7$.}
\begin{tabular}{lccccc}
\hline
 & \multicolumn{5}{c}{Soft X-ray band} \\
 & \multicolumn{5}{c}{[0.5--2] keV} \\
 & \multicolumn{2}{c}{Sample $A$} & & \multicolumn{2}{c}{Sample $B$} \\
\hline
 & $b_{\rm min}$ & $b_{\rm max}$ & & $b_{\rm min}$ & $b_{\rm max}$    \\
& \\
\ovisc & $3.45\pm0.39$ & $5.67\pm0.67$ && $2.82\pm0.38$ & $7.21\pm1.43$ \\
\lvisc & $3.36\pm0.33$ & $5.73\pm0.80$ && $2.73\pm0.79$ & $7.72\pm1.93$ \\
\csf   & $3.49\pm0.21$ & $5.33\pm0.57$ && $2.50\pm0.28$ & $7.14\pm1.14$ \\
\csfc  & $2.76\pm0.28$ & $6.36\pm1.09$ && $1.99\pm0.50$ & $7.80\pm1.76$ \\
\hline
 & \multicolumn{5}{c}{Hard X-ray band} \\
 & \multicolumn{5}{c}{[2--10] keV} \\
 & \multicolumn{2}{c}{Sample $A$} & & \multicolumn{2}{c}{Sample $B$} \\
\hline
 & $b_{\rm min}$ & $b_{\rm max}$ & & $b_{\rm min}$ & $b_{\rm max}$    \\
& \\
\ovisc & $3.10\pm0.55$ & $7.97\pm1.01$ && $3.42\pm0.71$ & $9.82\pm2.50$ \\
\lvisc & $3.08\pm0.29$ & $7.87\pm0.90$ && $3.36\pm1.72$ &$10.48\pm5.26$ \\
\csf   & $3.25\pm0.41$ & $7.56\pm0.73$ && $3.25\pm0.43$ & $9.69\pm2.66$ \\
\csfc  & $1.94\pm0.36$ & $9.31\pm2.11$ && $2.87\pm1.71$ &$10.95\pm5.79$ \\
\hline
\end{tabular}
\label{tab:fit_rpl}
\end{center}
\end{table*}

Since we observe a steepening of the slope of the X-ray profiles going to the external regions,
we fit them with a rolling-index power law function 
\begin{equation} 
y = ax^{-b_{\rm max}\frac{x+b_{\rm min}/b_{\rm max}}{x+1}} \ ,
\label{eq:roll_pow_law}
\end{equation}
where $x \equiv r/R_{200}$, and $a$, $b_{\rm  min}$ and $b_{\rm max}$ are
the free parameters, with $b_{\rm min}$ and $b_{\rm max}$ representing
the slope in the two asymptotic cases, $x \rightarrow 0$ and $x \gg 1$, respectively.
We fit the two samples separately in the interval $0.3 \leq r/R_{\rm 200} \leq 2.7$.
The results for the 4 physical models are quoted in Tab.~\ref{tab:fit_rpl}. The average  
best-fit values on the slopes of the surface brightness of the objects 
simulated with the \ovisc\ model are shown in Fig.~\ref{fig:fit_rpl}.  
Overall, the fits well reproduce the shape of the surface brightness
in the interval $0.3 \leq r/R_{200} \leq 2.7$. 
Our sample of galaxy ``groups" (sample $B$) presents a soft X-ray profile that
is (i) shallower in the centre ($b_{\rm min, B} \sim 2.7$,
$b_{\rm min, A} \sim 3.4$) and (ii) steeper in the outskirts
($b_{\rm max, B} \sim 7.3$, $b_{\rm max, A} \sim 5.5$)
than the one estimated in the sample of massive clusters (sample $A$).
The following reasons can be considered to explain this behaviour: 
(i) the ICM temperature of the objects in sample $B$ at $R_{200}$
is below 1 keV (see Table \ref{tab:9clusters}) and decreases at $r>R_{200}$ to values 
that becomes comparable to (and less than) the lower end of the [0.5--2] keV
band here considered, resulting in a cut-off of the soft X-ray signal.   
This also explains the steeper slope of the hard X-ray profiles
in the less massive objects;
(ii) the higher cooling efficiency in groups, with respect to rich clusters, enhances 
the selective removal of low-entropy gas from the hot phase, thus producing shallower 
profiles of X-ray emitting gas;
(iii) as discussed at the end of Section \ref{sect:simul}, while our clusters are the most massive 
objects in the volume of the parent simulation and are still accreting material,
the ``groups" appear to be isolated, thus with less evidence of recent accretion activity.
This results in slightly shallower profiles near the centre.

\subsection{Comparison with observational constraints} \label{sect:comparison}

\begin{table*}
\begin{center}
\caption{Distribution of the best-fit values of the power-law slopes of the X-ray
surface brightness profiles separated between the two samples $A$ and $B$ ($b_A$
and $b_B$, respectively), the two X-ray bands and different physical models. The
profiles extracted from the 99 per cent volume-selection scheme and fitted over
the different radial ranges $x_{\rm min} - x_{\rm max}$ (in units of $R_{200}$.)
Values of $b_{\rm obs}$ are the observational constraints in the same radial
range from \protect \cite{neumann2005} (see text in Section \ref{sect:comparison}).
}
\begin{tabular}{lccccccc}
\hline
 & & \multicolumn{3}{c}{Soft X-ray band} && \multicolumn{2}{c}{Hard X-ray band} \\
 & & \multicolumn{3}{c}{[0.5--2] keV} && \multicolumn{2}{c}{[2--10] keV} \\
\hline
 & $x_{\rm min} - x_{\rm max}$ & $b_A$ & $b_B$ & $b_{\rm obs}$ & & $b_A$ & $b_B$    \\
& \\
\ovisc &  0.3 -- 1.2  & 4.05$\pm$0.03 & 4.09$\pm$0.01 & 3.79$\pm$0.38 && 4.46$\pm$0.05 & 5.23$\pm$0.03 \\
\lvisc &  0.3 -- 1.2  & 4.04$\pm$0.04 & 4.14$\pm$0.09 &               && 4.47$\pm$0.05 & 5.32$\pm$0.17 \\
\csf   &  0.3 -- 1.2  & 3.99$\pm$0.00 & 3.81$\pm$0.02 &               && 4.50$\pm$0.02 & 5.07$\pm$0.04 \\
\csfc  &  0.3 -- 1.2  & 3.79$\pm$0.02 & 3.57$\pm$0.04 &               && 4.19$\pm$0.03 & 5.03$\pm$0.12 \\
\\
\ovisc &  0.7 -- 1.2  & 4.49$\pm$0.19 & 4.76$\pm$0.32 & 5.73$^{+1.43}_{-1.26}$ && 5.15$\pm$0.41 & 6.42$\pm$0.76 \\
\lvisc &  0.7 -- 1.2  & 4.54$\pm$0.32 & 4.66$\pm$0.19 &               && 5.26$\pm$0.44 & 6.20$\pm$0.75 \\
\csf   &  0.7 -- 1.2  & 4.31$\pm$0.28 & 4.45$\pm$0.11 &               && 5.18$\pm$0.37 & 5.92$\pm$0.13 \\
\csfc  &  0.7 -- 1.2  & 4.29$\pm$0.17 & 4.35$\pm$0.14 &               && 4.89$\pm$0.24 & 6.42$\pm$0.61 \\
\\
\ovisc &  1.2 -- 2.7  & 5.15$\pm$0.62 & 6.37$\pm$0.93 &               && 6.95$\pm$0.88 & 8.43$\pm$1.70 \\
\lvisc &  1.2 -- 2.7  & 5.21$\pm$0.59 & 6.77$\pm$0.95 &               && 6.88$\pm$0.54 & 9.01$\pm$2.72 \\
\csf   &  1.2 -- 2.7  & 4.90$\pm$0.53 & 6.23$\pm$0.67 &               && 6.70$\pm$0.74 & 8.40$\pm$1.52 \\
\csfc  &  1.2 -- 2.7  & 5.67$\pm$0.63 & 6.58$\pm$0.88 &               && 8.14$\pm$1.32 & 9.14$\pm$3.36 \\
\hline
\end{tabular}
\label{tab:fit_intervals}
\end{center}
\end{table*}

Our results on the external shape of the gas density, temperature and surface
brightness profiles in simulated galaxy clusters can be compared with recent estimates
obtained for nearby X-ray bright objects.
\cite{neumann2005} discusses the outer slope of the soft X-ray surface brightness
in 14 clusters observed with {\it ROSAT/PSPC}. In order to compare our results with 
the ones of that work, we fit the 99 per cent volume
X-ray profiles in different intervals for the two clusters samples with a single
power-law function, $S(x) = a \, x^{-b}$,
where $a$ and $b$ are the two free parameters and $x \equiv r/ R_{200}$.
We avoid to consider the interval $0.1 \leq x \leq 0.3$, owing to the fact that
this region is very model-dependent and cannot be represented by a single
power-law relation.
Our results are listed in Table \ref{tab:fit_intervals} together with the 
corresponding results ($b_{\rm obs}$) of \cite{neumann2005}. In the interval $0.3 \leq x 
\leq 1.2$, for the 
sample of clusters labeled "4" that better matches the virial temperatures of our sample $A$,
\cite{neumann2005} measures a slope of $3.79 \pm 0.38$ in good 
agreement with our average values of $(3.78, 4.03)$, being the lower values
measured in objects simulated with the inclusion of extra-physics 
(models \csf\ and \csfc).
In the interval $0.7 \leq x \leq 1.2$, the observational constraints are looser 
and are between $5.73^{+1.43}_{-1.26}$ for the 7 most massive clusters in the
Neumann's sample and $7.22^{+1.79}_{-1.60}$ for the whole dataset.
Our constraint of $\approx 4.3-4.5$ lays on the lower end of this distribution, but
still just $1 \sigma$ away from the results obtained for the hottest sample. 
The sample $B$ shows profiles with mean slopes consistent with those obtained
in sample $A$ within the measured dispersion (see Table~\ref{tab:fit_intervals}). 

For what concerns the temperature profile, a comparison can be done with the
functional form that reproduces the behaviour of the deprojected X-ray temperature 
profile at $r \ga 0.05 R_{200}$ of 13 low-redshift clusters observed with {\it Chandra} 
as presented in \citet[ see their equation~9]{vikhlinin2006}:
\begin{equation}
T(d) \propto \frac{ (d/0.045)^{1.9} +0.45}{(d/0.045)^{1.9} +1} \cdot
\frac{1}{(1+(d/0.6)^2)^{0.45}},
\end{equation} 
where $d \equiv r/ R_{500}$.
We overplot this function normalized to the average temperature value at 
$0.5 R_{200}$ to our profiles in Fig.~\ref{fig:compare}.

While it is known that in general hydrodynamical simulations cannot reproduce the steepening 
of the observed temperature profiles at the centre \citep[see][for more detail]{borgani2004,
borgani2006}, we can see that the agreement in the external regions is good for the most 
massive clusters, as already noted by \cite{markevitch1998} and \cite{degrandi2002}. 
The non-radiative models show a good agreement with this form, so that when 
we normalize the function at $0.5 R_{200}$ the difference between the values quoted in 
Table \ref{tab:prof_values} is less than $1 \sigma$ up to $2 R_{200}$ for both clusters 
and groups. We only find small deviations from the observed profile that become more 
significant at $r < 0.3 R_{200}$ for simulations including cooling and at $r > R_{200}$ 
where non-thermalized accreting material is dominant. The latter effect is particularly 
evident in low-mass systems.

\subsection{Implications on X-ray properties of the cluster virial regions}

Using our profiles we can predict the steepness of the profiles in the external regions 
of galaxy clusters. We fit the soft (0.5--2 keV) and hard (2--10 keV) X-ray profiles also in 
the interval $1.2 \leq r/R_{200} \leq 2.7$. The results are 
shown in Table \ref{tab:fit_intervals}. Our profiles are steeper in the external regions 
for all the models and for both the cluster samples: the steepening is more evident in 
sample $B$ suggesting a break of self-similarity at the scales around $2 R_{\rm 200}$. 
As already noted at the end of Section~\ref{sect:results}, this is mainly due to the fact 
that the temperature in the external regions of the haloes of sample $B$ drops below 
0.5 keV resulting in a cutoff of the soft X-ray signal.
From the fit with a rolling power-law, the values of $b_{\rm max}$
can be associated to the asymptotic behaviour at $r \gg R_{200}$.
We measure a slope of the surface brightness that ranges from
about $\sim7.5$ in groups to $\sim5.8$ in clusters.
In terms of the $\beta$ value of the $\beta-$model \citep[e.g.][]{cavaliere1978}, 
$\beta \approx (1 +b_{\rm max}) / 6$,
it approaches an estimate between $1.4$ in groups and $1.1$ in clusters, the latter
being a 40 per cent steeper than the observed values at $\sim R_{200}$ 
\citep{vikhlinin1999,neumann2005}.

\section{Summary and conclusions} \label{sect:concl}

Using a set of hydrodynamical simulations, performed with the Tree+SPH code 
{\tt GADGET-2}, composed by 9 galaxy clusters covering the mass range 
$1.5 \times 10^{14} M_{\odot} < M_{\rm vir} < 3.4 \times 10^{15} M_{\odot}$ and 
adopting 4 different physical prescriptions with fixed metallicity,
we have studied the behaviour in the outer regions of the gas density, temperature, 
soft (0.5--2 keV) and hard (2--10 keV) X-ray surface brightness profiles paying 
particular attention to the logarithmic slope.
Outside the central regions, where a not-well defined heating source
can partially or totally balance the radiative processes, the physics
of the X-ray emitting intracluster plasma is expected to be mainly driven
by adiabatic compression and shocks that take place during the
collapse of the cosmic baryons into accreting dark matter halos.
These processes are properly considered in cosmological numerical simulations 
here analyzed, allowing us to investigate on solid basis the expected behaviour
of the cluster outskirts. 

Our main findings can be summarized as follows\footnote{
Note that, as in the rest of the paper, 
the values of the slopes here quoted are in fact negative slopes.}:

\begin{enumerate}

\item the behaviour of the profiles in the external regions of clusters 
($r \gtrsim R_{200}$) does not depend significantly on the presence or absence of 
cooling and SN feedback. Only the inclusion of a model of thermal conduction 
can introduce non-negligible changes in the values of the slopes of the profiles.
This shows that the mean behaviour of the X-ray emitting plasma in the
cluster outskirts is mainly due to the gravitational force. 

\item The gas density profile steepens in the outskirts changing the slope
of the power-law from $\sim 2.5$ to $\sim 3.4$ at about $1.2 R_{200}$.
The temperature profile varies the slope of the power-law like functional 
form from $ 0.4$ in non-radiative models and $ 0.6$ in models with cooling and
feedback, to about $ (1.7-1.9)$ at $\sim1.3-1.6 R_{200}$.

\item The surface brightness in the outskirts profiles appear to be dominated by the
presence of subclumps of cold and dense material that we exclude from our analyses.
After removing these objects, the surface brightness in the [0.5--2] keV presents a
profile with a slope that steepens towards the external regions. Fitting the profiles with
a power-law function in different radial ranges between 0.3 and 2.7 $R_{200}$ we
obtain slopes between 4 and 5.5 in the 4 most massive systems and between $ 3.8$ and $ 6.5$
in the 5 objects with masses $\approx 10^{14} M_{\odot}$.

\item We introduce a new rolling-index power-law functional (eq. \ref{eq:roll_pow_law})
which fits well the X-ray surface profiles in the interval $0.3 \leq r/R_{200} \leq 2.7$.
From the results of these fits, we obtain that our sample of groups of galaxies (sample $B$)
has soft X-ray profiles shallower in the centre ($b_{\rm min, B} \sim 2.6$, 
$b_{\rm min, A} \sim  3.3$) and steeper in the outskirts
($b_{\rm max, B} \sim 7.4$, $b_{\rm max, A} \sim 5.8$)
than our sample of clusters (sample $A$).
This is mainly an effect of the drop out from the pass band of the 
soft X-ray emission, due to the low ICM temperature of the objects in sample $B$ at the 
virial radius.

\item We compare our results on the shape of the temperature and soft X-ray surface brightness
profiles with the present observational constraints.
By using the functional form of \cite{vikhlinin2006} (eq. 9) extrapolated
beyond $R_{500}$, we verify that there is a good agreement with the simulated profiles
over the radial range $0.3 - 2 R_{200}$.
On the surface brightness profile in the [0.5--2] keV band, we measure a slope
in the outskirts that is consistent within $1 \sigma$ with the values estimated
from {\it ROSAT/PSPC} data in \cite{neumann2005}.

\item From the overall shape of the simulated profiles, we predict that 
(see Table~\ref{tab:prof_values}) 
(1) the ICM temperature profile decreases with radius reaching at 
$R_{200}$ average values of $0.60$ times the value measured at $0.3 \times R_{200}$,
(2) the estimated gas density at $R_{200}$
is $4.8 \times 10^{-3}$ times the gas density measured at $0.3 R_{200}$, 
(3) the surface brightness in the soft band
is about $10^{-12}$ erg s$^{-1}$ cm$^{-2}$ deg$^{-2}$ at $R_{200}$
at $z$=0, that is a factor of few below the extragalactic unresolved X-ray 
background \citep[see][]{hickox2006}.

\end{enumerate}

This last item shows that the perspective to resolve the X-ray 
emission around the cluster virial regions is extremely challenging 
from an observational point of view, even at low redshift, 
unless tight constraints on all the components of the 
measured background can be reached \citep[see also][]{molendi2004}. 
Considering the similarity of the surface brightness profiles 
of the objects of the same mass, the observational technique of
stacking the images of different clusters \citep[e.g.][]{neumann2005}
can help in enhancing the statistics of the profiles in the 
outer regions.

Overall, our results show how the study of simulated objects 
can give indications on the requested capabilities of future 
experiments in order to resolve the cluster virial regions. On the 
other hand, since the behaviour of the slopes have shown to be mainly 
independent of the physical models adopted in our simulations, any 
discrepancy between our results and observations of the clusters 
virial regions could provide useful hints on the presence of additional 
physics (e.g. turbulence, magnetic fields, cosmic rays) influencing 
the thermodynamics of the ICM.

Finally, the work presented here further demonstrates the 
complementarity between numerical and observational cosmology in 
describing the properties of highly non-linear and over dense structures.

\section*{acknowledgements}

The simulations were carried out on the IBM-SP4 and IBM-SP5 machines 
and on the IBM-Linux cluster at the ``Centro Interuniversitario del 
Nord-Est per il Calcolo Elettronico'' (CINECA, Bologna), with CPU 
time assigned under INAF/CINECA and
University-of-Trieste/CINECA grants, on the IBM-SP3 at the Italian Centre of
Excellence ``Science and Applications of Advanced Computational Paradigms'',
Padova and on the IBM-SP4 machine at the ``Rechenzentrum der
Max-Planck-Gesellschaft'', Garching. This work has been partially 
supported by the PD-51 INFN grant. We acknowledge financial contribution from 
contract ASI-INAF I/023/05/0. We wish to thank the anonymous 
referee for useful comments that improved the presentations of our 
results. We are also grateful to F. Civano, A. Morandi, E. Rasia, 
C. Tonini and F. Vazza for useful discussions.

\appendix

\section{Effects of volume cutting in density and temperature profiles}
\label{app:volume_cut}

In this appendix we will briefly discuss the effects of the volume--selection scheme 
adopted in this paper.
We show in Table \ref{tab:fit_dens} and \ref{tab:fit_temp} the results for the fits 
of the density and temperature profiles calculated with different cuts in volume 
using a broken power-law function of eq. \ref{eq:broken_pow_law} in the logarithmic form.

For the density profiles the internal slope does not depend much on the physics or the 
volume cutting scheme adopted: this shows that our method does not introduce any 
bias to the results of density for the internal regions of clusters.
The external slope for the density profiles does not show any significant dependence on 
the physics but it decreases regularly when we include more and more cool and diffuse gas: 
this indicates that to obtain a fair representation of the cluster temperature in the 
outskirts it is necessary to consider the 99 per cent volume profiles. If we include all 
the gas particles the value of $b_2$ continues to go down, but the profiles begin to show 
features due to the cold and gas and the $\chi^2$ of the fit increases. For this 
reason and since the 99 per cent profiles show a remarkable regularity, as already said 
in the comments to Fig. \ref{fig:profiles}, we can assume these as representative of the 
global cluster properties, although there may be some small systematic uncertainties 
due to our method.

As already discussed in Section~\ref{sect:results}, the presence of cooling in the simulations 
creates steeper temperature profiles at the centre, and this can be seen by the different 
values of $b_1$ in Table \ref{tab:fit_temp}. Again, like for the temperature profiles, 
the value of the internal slope is not influenced by the volume selection scheme.
On the contrary the values of $b_2$ are much less affected by the physics but show an 
increasing trend when adding colder gas, considering also a significant scatter between 
the 9 clusters. The difference between the 99 and 100 per cent values for $b_2$ is 
particularly evident with the \ovisc\ model, but, again, given the higher values of the 
$\chi^2$ and the regularity of the 99 per cent profiles, we can draw the same conclusions 
as for the density profiles.

\begin{table*}
\begin{center}
\caption{
Best-fit results for the inner slope $b_1$, outer slope $b_2$ and break radius $R_b$ of the gas density 
profiles fitted with a broken power-law. The profiles are computed considering different cuts in volume, 
from 50 to 100 per cent.
}
\begin{tabular}{lcccccc}
Volume cut  &        50\%       &       80\%        &       90\%        &       95\%        &       99\%        &     100\%  \\
\hline

 & \multicolumn{6}{c}{$b_1$}                               \\

\ovisc      &  $2.47 \pm 0.03$  &  $2.50 \pm 0.03$  &  $2.50 \pm 0.03$  &  $2.47 \pm 0.03$  &  $2.46 \pm 0.03$  &  $2.55 \pm 0.02$  \\
\csf        &  $2.42 \pm 0.02$  &  $2.45 \pm 0.01$  &  $2.43 \pm 0.02$  &  $2.37 \pm 0.02$  &  $2.46 \pm 0.03$  &  $2.48 \pm 0.03$  \\
\hline

 & \multicolumn{6}{c}{$b_2$}                               \\

\ovisc      &  $3.86 \pm 0.55$  &  $3.78 \pm 0.43$  &  $3.71 \pm 0.39$  &  $3.60 \pm 0.37$  &  $3.38 \pm 0.29$  &  $3.12 \pm 0.32$  \\
\csf        &  $3.85 \pm 0.45$  &  $3.76 \pm 0.37$  &  $3.66 \pm 0.33$  &  $3.53 \pm 0.32$  &  $3.41 \pm 0.19$  &  $2.98 \pm 0.38$  \\
\hline

 & \multicolumn{6}{c}{$R_b/R_{200}$}                                   \\

\ovisc      &  $1.12 \pm 0.14$  &  $1.20 \pm 0.14$  &  $1.20 \pm 0.15$  &  $1.14 \pm 0.15$  &  $1.14 \pm 0.20$  &  $1.23 \pm 0.14$ \\
\csf        &  $1.15 \pm 0.14$  &  $1.24 \pm 0.12$  &  $1.23 \pm 0.13$  &  $1.10 \pm 0.16$  &  $1.31 \pm 0.28$  &  $1.34 \pm 0.15$ \\
\hline

\end{tabular}
\label{tab:fit_dens}
\end{center}
\end{table*}

\begin{table*}
\begin{center}
\caption{
Like Table \ref{tab:fit_dens} but for the mass-weighted temperature.
}
\begin{tabular}{lcccccc}
Volume cut  &        50\%       &        80\%       &        90\%       &        95\%       &        99\%      &     100\%  \\
\hline

 & \multicolumn{6}{c}{$b_1$}                               \\
\ovisc      & $0.43 \pm 0.03$ & $0.43 \pm 0.02$ & $0.47 \pm 0.02$ & $0.48 \pm 0.02$ & $0.41 \pm 0.02$  &  $0.54 \pm 0.03$  \\
\csf        & $0.47 \pm 0.05$ & $0.55 \pm 0.02$ & $0.55 \pm 0.02$ & $0.55 \pm 0.02$ & $0.57 \pm 0.02$  &  $0.59 \pm 0.02$  \\

\hline

 & \multicolumn{6}{c}{$b_2$}                               \\
\ovisc      & $1.30 \pm 0.63$ & $1.22 \pm 0.53$ & $1.31 \pm 0.54$ & $1.49 \pm 0.51$ & $1.73 \pm 0.55$  &  $2.15 \pm 0.65$  \\
\csf        & $1.35 \pm 0.67$ & $1.38 \pm 0.51$ & $1.46 \pm 0.47$ & $1.62 \pm 0.49$ & $1.90 \pm 0.46$  &  $2.07 \pm 0.64$  \\
\hline

 & \multicolumn{6}{c}{$R_b/R_{200}$}                                   \\
\ovisc      & $1.21 \pm 0.24$ & $1.19 \pm 0.25$ & $1.33 \pm 0.21$ & $1.40 \pm 0.17$ & $1.29 \pm 0.24$  &  $1.49 \pm 0.18$  \\
\csf        & $1.20 \pm 0.24$ & $1.44 \pm 0.22$ & $1.46 \pm 0.20$ & $1.48 \pm 0.18$ & $1.56 \pm 0.18$  &  $1.50 \pm 0.17$  \\
\hline
\end{tabular}
\label{tab:fit_temp}
\end{center}
\end{table*}

\label{lastpage}
\end{document}